\documentclass[subeqn,12pt, a4paper]{article} 

\usepackage{amssymb,amsmath,amsfonts,amsthm, amscd, mathrsfs,helvet, mathtools}
\usepackage{bm, stmaryrd}
\usepackage{graphicx,verbatim}
\usepackage[usenames, dvipsnames]{color}
\usepackage{psfrag}
\usepackage[all]{xy}
\usepackage{tikz}
\usetikzlibrary{arrows,matrix,decorations.markings,shapes.geometric}
\usepackage[normalem]{ulem}

\theoremstyle{plain}

\newtheorem*{theorem*}{Theorem}

\newtheorem*{proposition*}{Proposition}

\newcommand{\tensor}[1]{{\bf \underline{#1}}}

\definecolor{brightBlue}{rgb}{0,0,1}

\definecolor{Violet}{rgb}{0.47,0,1}

 \DeclareMathOperator{\str}{str}

\def\b{\mathfrak{b}}

\def\f{\mathfrak{f}}
\def\g{\mathfrak{g}}
\def\h{\mathfrak{h}}

\def\n{\mathfrak{n}}
\def\p{\mathfrak{p}}

\def\Gr{\mathcal{G}r}

\def\ha{\mbox{\small $\frac{1}{2}$}}
\def\qa{\mbox{\small $\frac{1}{4}$}}

\def\C{\mathcal{C}}
\def\CC{\mathbb{C}}

\def\R{\mathcal{R}}
\def\L{\mathcal{L}}
\def\M{\mathcal{M}}

\def\1{\tensor{1}}
\def\2{\tensor{2}}
\def\3{\tensor{3}}
\def\4{\tensor{4}}

\def\beq{\begin{equation}}
\def\eeq{\end{equation}}
\def\beqz{\begin{equation*}}
\def\eeqz{\end{equation*}}
\def\bea{\begin{eqnarray}}
\def\eea{\end{eqnarray}}
\def\dd{d}
\def\dt{\widetilde{d}}
\def\JJ{J}
\def\Jt{\widetilde{J}}

\numberwithin{equation}{section}

\begin{document}

\begin{center}
\vspace*{2em}
{\large\bf Derivation of the action and symmetries \\[1.5mm]
of the $q$-deformed $AdS_5 \times S^5$ superstring}\\
\vspace{1.5em}
F. Delduc$\,{}^1$, M. Magro$\,{}^1$, B. Vicedo$\,{}^2$

\vspace{1em}
\begingroup\itshape
{\it 1) Laboratoire de Physique, ENS Lyon
et CNRS UMR 5672, Universit\'e de Lyon,}\\
{\it 46, all\'ee d'Italie, 69364 LYON Cedex 07, France}\\
\vspace{1em}
{\it 2) School of Physics, Astronomy and Mathematics,
University of Hertfordshire,}\\
{\it College Lane,
Hatfield AL10 9AB,
United Kingdom}
\par\endgroup
\vspace{1em}
\begingroup\ttfamily
Francois.Delduc@ens-lyon.fr, Marc.Magro@ens-lyon.fr, Benoit.Vicedo@gmail.com
\par\endgroup
\vspace{1.5em}
\end{center}

\paragraph{Abstract.}

We recently proposed an integrable $q$-deformation of
the $AdS_5 \times S^5$ superstring action. Here we give details on
the hamiltonian origin and construction of this deformation. The procedure is a
generalization of the one previously developed for deforming principal chiral and
symmetric space $\sigma$-models. We also show that the original
$\mathfrak{psu}(2,2|4)$ symmetry is replaced in the deformed theory
by a classical analog of the quantum group $U_q(\mathfrak{psu}
(2,2|4))$ with $q$ real. The relation between $q$ and the deformation
parameter $\eta$ entering the action is given. The framework used to derive the
deformation also enables to prove that at the hamiltonian level,
the ``maximal deformation'' limit corresponds to an undeformed
semi-symmetric space $\sigma$-model with bosonic part $dS_5 \times H^5$.
Finally, we discuss the various freedoms in the construction.

\setcounter{tocdepth}{2}

\section{Introduction}

In \cite{Delduc:2013fga} we presented a general method
for constructing classical integrable deformations of principal chiral and symmetric space $\sigma$-models. At the hamiltonian level, the classical integrability of these $\sigma$-models rests on the fact that the Poisson bracket of their Lax matrix takes the general form in \cite{Maillet:1985fn,Maillet:1985ek}.
An important related feature of these $\sigma$-models is the existence of
another compatible Poisson bracket with respect to which the integrable
structure may be described \cite{Delduc:2012qb}.
The deformation is set up by starting from a linear combination of these
compatible Poisson brackets.
The same procedure may also be applied to the $AdS_5
\times S^5$ superstring. Indeed, it is known that the
Poisson bracket of the
corresponding Lax matrix has the right form
\cite{Magro:2008dv,Vicedo:2009sn}. Furthermore, the second compatible
Poisson bracket was obtained in \cite{Delduc:2012mk,Delduc:2012vq}.

\medskip

The deformed action in the case of the $AdS_5 \times S^5$ superstring was presented
in the letter \cite{Delduc:2013qra} where its classical integrability and $\kappa$-symmetry
invariance were also exhibited.
The action depends on a real
parameter $\eta \in \left] -1,1 \right[$ with $\eta =0$ corresponding to the undeformed
Metsaev-Tseytlin action \cite{Metsaev:1998it}.
The first purpose of this article is to present a derivation of
this deformed action within the hamiltonian framework. In fact, the latter is
also the right framework for studying
how the original $\mathfrak{psu}(2,2|4)$
symmetry is affected by the deformation.
The second purpose of this article is to show that this symmetry
gets replaced in the deformed theory by
the classical analog of $U_q(\mathfrak{psu}(2,2|4))$, where the
relation between $q$ and $\eta$ is found to be  
\begin{equation*}
q = \exp \left( - \frac{2 \eta (1 - \eta^2)}{(1 + \eta^2)^2} \right).
\end{equation*}
This relation, which may in fact
already be inferred from the bosonic case \cite{Delduc:2013fga},
is in agreement with the one found in
\cite{Arutyunov:2013ega}. We also indicate why the limits $\eta \to \pm 1$
correspond, at the hamiltonian level, to an undeformed
semi-symmetric space $\sigma$-model. In particular, we show that its
target space is $PSU^\ast(4|4)/(SO(4,1) \times SO(5))$, the
bosonic sector of which corresponds to $dS_5 \times H^5$. This proves the
conjecture made in \cite{Delduc:2013qra}.

\medskip

We discuss the various freedoms and rigidities in the construction.
The linear combination
of the two Poisson brackets used in defining the deformation is characterised
by a so called deformed twist function. We argue that this function is essentially unique.
This rules out the possibility of obtaining a double deformation
within this framework. On the other hand, another key ingredient in the construction
is a so called non-split $R$-matrix on $\mathfrak{psu}(2,2|4)$. We discuss what happens
when one considers other non-split $R$-matrices than the one considered in
\cite{Delduc:2013qra}.

\medskip

The plan of the article is the following. In section \ref{sec_GS}, we recall
important properties related to the hamiltonian integrability of the Green-Schwarz
superstring on $AdS_5 \times S^5$. The deformation is then carried out
at the hamiltonian level in section
\ref{sec_defo}. The limits $\eta \to \pm 1$ are discussed in subsection \ref{ds5h5}.
We show in section \ref{sec: qdefsymmetry} how the original
$\mathfrak{psu}(2,2|4)$ symmetry becomes $q$-deformed.
In section \ref{sec_action}, we perform the inverse Legendre transform
to determine the deformed action, which was presented in the letter
\cite{Delduc:2013qra}.
Some open questions are mentioned in the conclusion.
This article contains four appendices. Properties of
$\mathfrak{psu}(2,2|4)$ which are used have been
collected in
appendix \ref{app: psu}. Appendix  \ref{app: nsrmatrix}
concerns  non-split $R$-matrices.
The $q$-Poisson-Serre relations are proved in appendix
\ref{app: qSerre}. Finally, the discussion related to the choice
of $R$ is presented in appendix \ref{sec_invR}. In particular, we give the metrics and $B$-fields corresponding to three inequivalent choices of $R$-matrices in the case of $\mathfrak{su}(2,2)$.

\section{Green-Schwarz superstring on $AdS_5 \times S^5$} \label{sec_GS}

We start this section by recalling properties of the hamiltonian integrability
of the $AdS_5 \times S^5$ superstring that will be used. For more details
concerning material  presented in subsections \ref{subsec 21}
and \ref{subsec 22}, see \cite{Vicedo:2009sn,Vicedo:2010qd,Magro:2010jx}.

\subsection{Poisson bracket and Hamiltonian}
\label{subsec 21}
 To fix notations,  consider the real Lie superalgebra $\mathfrak{su}(2,2|4)$ and define the
Lie algebra $\f$ as its Grassmann envelope. We equip $\f$ with
a $\mathbb{Z}_4$-automorphism $\Omega$ defined in equation \eqref{Omega def}.
The corresponding decomposition of $\f$ into the eigenspaces of $\Omega$ is
 $\f = \f^{(0)} \oplus \f^{(1)} \oplus \f^{(2)} \oplus \f^{(3)}$. Define the Lie group $F = \exp \f$
and the subgroup $G = \exp \g$ associated with the Lie subalgebra
$\g = \f^{(0)}$. We refer the reader to appendix \ref{app: psu} for further details.

At the hamiltonian level, the supersymmetric $\sigma$-model on the
semi-symmetric space $F/G$
 may be described by a pair of fields
$A$ and $\Pi$ taking values in the Lie algebra $\f$. We shall
consider the case where the underlying space, parameterised
by $\sigma$, is the entire real line. The fields $A$ and $\Pi$,
which are assumed to decay sufficiently rapidly at infinity,
satisfy the following Poisson brackets
\begin{subequations} \label{PB string}
\begin{align}
\label{PB string a} \big\{ A^{(i)}_{\1}(\sigma), A^{(j)}_{\2}(\sigma') \big\} &= 0,\\
\label{PB string b} \big\{ A^{(i)}_{\1}(\sigma), \Pi^{(j)}_{\2}(\sigma') \big\} &= \big[ C_{\1\2}^{(i \, 4 - i)}, A^{(i+j)}_{\2}(\sigma) \big] \delta_{\sigma \sigma'} - C_{\1\2}^{(i \, 4 - i)} \delta_{i+j, 0} \, \partial_{\sigma} \delta_{\sigma \sigma'},\\
\label{PB string c} \big\{ \Pi^{(i)}_{\1}(\sigma), \Pi^{(j)}_{\2}(\sigma') \big\} &=
 \big[ C_{\1\2}^{(i \, 4 - i)}, \Pi^{(i+j)}_{\2}(\sigma) \big] \delta_{\sigma \sigma'}.
\end{align}
\end{subequations}
Here $C_{\1\2}^{(i \, 4 - i)}$ is the projection onto $\f^{(i)} \otimes \f^{(4-i)}$ of the    quadratic Casimir $C_{\1\2}$ defined by equation \eqref{psl Casimir}
 and $\delta_{\sigma \sigma'}=\delta(\sigma-\sigma')$ is
the Dirac distribution.
 There are also the following constraints:
\begin{subequations} \label{constraints}
\begin{align}
\label{constraints a} \C^{(0)} &= \Pi^{(0)} \simeq 0,\\
\label{constraints b} \C^{(1)} &= \ha A^{(1)} + \Pi^{(1)} \simeq 0,\\
\label{constraints c} \C^{(3)} &= - \ha A^{(3)} + \Pi^{(3)} \simeq 0,\\
\label{constraints d} T_{\pm} &= \str(A_{\pm}^{(2)} A_{\pm}^{(2)}) \simeq 0,
\end{align}
\end{subequations}
where
\beq \label{def Apm2}
A_{\pm}^{(2)} = \ha (\Pi^{(2)} \mp A^{(2)}).
\eeq
The constraint $\C^{(0)}$ is associated with the $SO(4,1) \times
SO(5)$ gauge invariance and $T_\pm$ are the Virasoro constraints.
 The fermionic constraints   $\C^{(1)}$
and $\C^{(3)}$ are a mixture of first-class and second-class
constraints. Their first-class part,
\begin{equation} \label{Kappa13}
\mathcal{K}^{(1)} = 2 i \big[ A^{(2)}_-, \C^{(1)} \big]_+, \qquad
\mathcal{K}^{(3)} = 2 i \big[ A^{(2)}_+, \C^{(3)} \big]_+,
\end{equation}
is related to the $\kappa$-symmetry of the superstring. We
introduce the following quantities:
\begin{equation} \label{Tpmpm}
\mathcal{T}_+ =  T_+ - \str \big( (A^{(1)} - \ha \C^{(1)}) \C^{(3)} \big), \qquad
\mathcal{T}_- =   T_- + \str \big( (A^{(3)} + \ha \C^{(3)}) \C^{(1)} \big).
\end{equation}
Then the dynamics is induced \cite{Vicedo:2009sn} by the Hamiltonian
$H_{\rm string} = \int_{- \infty}^{\infty} d\sigma h_{\rm string}$ where
\begin{equation} \label{Ham string}
h_{\rm string} = \lambda^+ \mathcal{T}_+
+ \lambda^- \mathcal{T}_- - \str(k^{(3)} \mathcal{K}^{(1)})
- \str(k^{(1)} \mathcal{K}^{(3)}) - \str\big( (A^{(0)} + \ell) \Pi^{(0)} \big).
\end{equation}
Here the variables $\lambda^{\pm}$ are related to the worldsheet metric $h_{\alpha \beta}$ as
\begin{equation} \label{deflambdapm}
\lambda^{\pm} = \frac{1 \pm \gamma_{01}}{\gamma_{11}} =
 \frac{1 \pm \gamma^{01}}{- \gamma^{00}},
\end{equation}
where $\gamma_{\alpha \beta} = \sqrt{- h} h_{\alpha \beta}$.

\subsection{Lax matrix and integrability}
\label{subsec 22}
The $AdS_5 \times S^5$ superstring possesses an infinite
number of hidden symmetries. In order to identify them
we rephrase the Poisson bracket \eqref{PB string} together
with the dynamics induced by the Hamiltonian \eqref{Ham string}
in terms of the so called Lax matrix. In the present case,
the latter is a linear combination of the
 fields  $(A,\Pi)$
and depends on an arbitrary complex variable $z$ called
the spectral parameter, namely
\begin{multline} \label{Lax matrix}
\L(z) = A^{(0)} + \qa (z^{-3} + 3 z) A^{(1)} + \ha (z^{-2} + z^2) A^{(2)} + \qa (3 z^{-1} + z^3) A^{(3)}\\
+ \ha (1 - z^4) \Pi^{(0)} + \ha (z^{-3} - z) \Pi^{(1)} + \ha (z^{-2} - z^2) \Pi^{(2)} + \ha (z^{-1} - z^3) \Pi^{(3)}.
\end{multline}
Its first property is that the Poisson brackets \eqref{PB string} of the
 fields $(A,\Pi)$ are satisfied if and only if the Poisson bracket
of the Lax matrix \eqref{Lax matrix} with itself takes the form
\begin{equation} \label{Lax PB}
\big\{ \L_{\1}(\sigma), \L_{\2}(\sigma') \big\} = \big[ \R_{\1\2}, \L_{\1}(\sigma) \big] \delta_{\sigma \sigma'} - \big[ \R^{\ast}_{\1\2}, \L_{\2}(\sigma) \big] \delta_{\sigma \sigma'} + \big( \R_{\1\2} + \R^{\ast}_{\1\2} \big) \partial_{\sigma} \delta_{\sigma \sigma'},
\end{equation}
where the notation is as follows.
We use the usual shorthands $\L_{\1} = \L(z_1) \otimes 1$, $\L_{\2} = 1 \otimes \L(z_2)$ where the dependence on the pair of spectral parameters $z_1$, $z_2$ is implicit in the tensorial index. Similarly, the $\R$-matrix, which lives in both tensor factors, also depends on both spectral parameters and is given explicitly by
\begin{equation} \label{R-matrix}
\R_{\1\2}(z_1, z_2) = 2 \frac{\sum_{j=0}^3 z_1^j z_2^{4-j} C_{\1\2}^{(j \, 4-j)}}{z_2^4 - z_1^4} \; \phi_{\rm string}(z_2)^{-1}, \qquad \phi_{\rm string}(z) = \frac{4 z^4}{(1 - z^4)^2}.
\end{equation}
We refer to $\phi_{\rm string}(z)$ as the twist function.
 The adjoint of the $\R$-matrix in \eqref{Lax PB} is then given simply by $\R^{\ast}_{\1\2}(z_1, z_2) = \R_{\2\1}(z_2, z_1)$.

Secondly, the evolution of the
  fields $A$ and $\Pi$
under the string Hamiltonian $H_{\rm string}$ is equivalent
to the following zero-curvature equation
\begin{equation} \label{ZC eq}
\big[ \partial_{\tau} - \M(z), \partial_{\sigma} - \L(z) \big] = 0,
\end{equation}
governing the time evolution of the Lax matrix \eqref{Lax matrix}.
Here
$\partial_{\tau} \equiv \{  \cdot, H_{\rm string} \}$ and we have introduced
\begin{multline} \label{time Lax matrix}
\M(z) = A^{(0)} - \qa (z^{-3} - 3 z) A^{(1)} - \ha (z^{-2} - z^2) A^{(2)} - \qa (3 z^{-1} - z^3) A^{(3)}\\
+ \ha (1 - z^4) \Pi^{(0)} - \ha (z^{-3} + z) \Pi^{(1)} - \ha (z^{-2} + z^2) \Pi^{(2)} - \ha (z^{-1} + z^3) \Pi^{(3)}.
\end{multline}

The advantage of formulating the Poisson structure and
dynamics of the superstring $\sigma$-model in the
Lax form \eqref{Lax PB} and \eqref{ZC eq} is that it naturally
 lends itself to the construction of an infinite number of
conserved charges in involution. Specifically, if we define
the monodromy matrix as
\begin{equation} \label{monodromy}
T(z) = P \overleftarrow{\exp} \int_{- \infty}^{\infty} d\sigma \L(z),
\end{equation}
then by the usual argument
 it follows directly from \eqref{ZC eq}  and the decay of the fields at infinity
that $T(z)$ is conserved, namely
\begin{equation} \label{T cons}
\partial_{\tau} T(z) = 0.
\end{equation}
Expanding the monodromy in $z - 1$ then yields an infinite number
of non-local conserved charges.

\subsection{Group valued field}

The group valued field $g$ of the semi-symmetric
space $\sigma$-model is defined in terms of $A$
through the relation $A = - g^{-1} \partial_{\sigma} g$. If we
also define the field $X = - g \Pi g^{-1}$ then the
  Poisson brackets \eqref{PB string} can be deduced from
\begin{subequations} \label{PB gX}
\begin{align}
\label{PB gX a} \{ g_{\1}(\sigma), g_{\2}(\sigma') \} &= 0,\\
\label{PB gX b} \{ X_{\1}(\sigma), g_{\2}(\sigma') \} &= C_{\1\2} \, g_{\2}(\sigma) \delta_{\sigma \sigma'},\\
\label{PB gX c} \{ X_{\1}(\sigma), X_{\2}(\sigma') \} &= \big[ C_{\1\2}, X_{\2}(\sigma) \big] \delta_{\sigma \sigma'}.
\end{align}
\end{subequations}

An important observation for what follows is the fact that the fields $g$ and $X$ can be obtained from the expansion of the Lax matrix \eqref{Lax matrix} at the poles of the twist function $\phi_{\rm string}(z)$. In order to see this, first note that the expansions at each of these four poles are related to one another using the relation $\Omega\big( \L(z) \big) = \L(i z)$. It is therefore sufficient to consider one of these poles, say $z = 1$.
Now the expansion of the Lax matrix near $z = 1$ reads
\beq
\L(z) = A - 2 (z - 1) \Pi + O\big( (z - 1)^2 \big). \label{Latzegal1}
\eeq
Consider the gauge transformation of the Lax matrix $\L^g(z) =
\partial_{\sigma} g g^{-1} + g \L(z) g^{-1}$ with the group valued
field $g$ as parameter. Using the relation $A = - g^{-1}
\partial_{\sigma} g$ we observe that
\begin{equation} \label{Lax expand}
\L^g(z) = 2 (z - 1) X + O\big( (z - 1)^2 \big).
\end{equation}
In particular, the group valued field $g$ is characterised
by the vanishing of the gauge transformed Lax matrix
$\L^g(z)$ at the special point $z = 1$. Furthermore,
the field $X$ corresponds to the
subleading term in the expansion of $\L^g(z)$ at that point.
In other words, we have
\begin{equation} \label{extract gX}
\L^g(1) = 0, \qquad
X = \ha \frac{d \L^g}{d z}(1).
\end{equation}

\subsection{Global symmetry algebra} \label{sec: sym undef}

We assume that the field $g$ tends to   constant values as
$\sigma \to \pm \infty$.   Then by virtue
of the conservation of the monodromy matrix \eqref{T cons}
it follows that its gauge transformation $T^g(z)$ by the field
$g$ is also conserved.  Using equation \eqref{Lax expand},
the first non-trivial terms in the expansion of the gauge
transformed monodromy near $z = 1$ read
\begin{equation} \label{Tg near 1}
T^g(z) = g(\infty) T(z) g(- \infty)^{-1} = P \overleftarrow{\exp}
\int_{-\infty}^{\infty} d\sigma \L^g(z) = {\bf 1} + 2 (z - 1)
\int_{-\infty}^{\infty} d\sigma X + O\big( (z-1)^2 \big).
\end{equation}
Since $T^g(z)$ is conserved for all $z$ it follows that
$\int_{-\infty}^{\infty} d\sigma X$ is conserved. It then
follows using \eqref{PB gX c} that its Poisson bracket algebra takes the form
\begin{equation*}
\left\{ \int_{-\infty}^{\infty} d\sigma X_{\1}, \int_{-\infty}^{\infty}
d\sigma X_{\2} \right\} = \left[ C_{\1\2}, \int_{-\infty}^{\infty} d\sigma X_{\2} \right].
\end{equation*}
This conserved charge therefore generates the symmetry under left action
by the Lie group $F$.

\section{Defining the deformation} \label{sec_defo}

We shall proceed to deform the $AdS_5 \times S^5$ superstring
$\sigma$-model by following the strategy developed in
\cite{Delduc:2013fga} for deforming symmetric space $\sigma$-models.
Thus, in order to preserve integrability throughout the deformation,
we shall not modify the Lax matrix \eqref{Lax matrix}. We shall also
not modify the dynamics of the fields
$A$ and $\Pi$.
In other words, the zero curvature equation \eqref{ZC eq} will remain
the same. All we will deform is the Poisson bracket \eqref{Lax PB}.
And in order to do so, we shall simply deform the twist function
appearing in the $\R$-matrix \eqref{R-matrix}, replacing
$\phi_{\rm string}$ by another function $\phi_{\epsilon}$ in such
a way that $\phi_{\epsilon} \to \phi_{\rm string}$ in the limit
$\epsilon \to 0$. It will then be a matter of suitably deforming
the relations \eqref{extract gX} for defining the fields $g$ and
$X$ of the deformed theory.

\subsection{The twist function}

Consider, therefore, the Poisson bracket \eqref{Lax PB} with the $\R$-matrix defined using a more general twist function, namely
\begin{equation} \label{R twist gen}
\R_{\1\2}(z_1, z_2) = 2 \frac{\sum_{j=0}^3 z_1^j z_2^{4-j} C_{\1\2}^{(j \, 4-j)}}{z_2^4 - z_1^4} \; \phi(z_2)^{-1}.
\end{equation}
With $\phi(z)$ set to $1$ this is simply the kernel of the standard $\R$-matrix
on the twisted loop algebra $\f^{\Omega}(\!(z)\!)$ with respect to the trigonometric
inner product.  More generally, the expression \eqref{R twist gen} is the kernel
of the same $\R$-matrix but with respect to a twisted inner product
(see for instance \cite{Vicedo:2010qd}). In order for the latter to
be non-degenerate
on $\f^{\Omega}(\!( z )\!)$, the twist function should satisfy $\phi(i z) = \phi(z)$.
On the other hand, it is well known \cite{Reyman:1988sf} that the
Poisson bracket
\eqref{Lax PB}
with $\R$-matrix \eqref{R twist gen} and twist function of the form
$\phi(z) = z^k$ leads to a well defined Poisson bracket for the fields
$A$ and $\Pi$ only if $- 4 \leq k \leq 4$.
 Hence there are only
three independent choices for the inverse of the twist function $\phi(z)^{-1}$ in
\eqref{R twist gen}, namely $z^4$, $1$ and $z^{-4}$. Moreover,
the corresponding brackets are all compatible
   \cite{Reyman:1988sf}. That is, any linear
combination of these also defines a valid Poisson bracket through
\eqref{Lax PB} and \eqref{R twist gen}.   Note that $\phi_{\rm string}(z)^{-1}
= \frac{1}{16} z^4 - \frac{1}{8} + \frac{1}{16} z^{-4}$ is such a linear combination.

\begin{figure}[h]
\centering
\def\svgwidth{125mm}
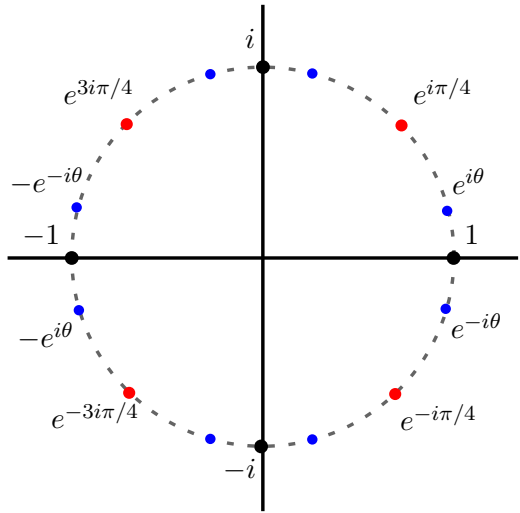
\caption{The eight poles of the deformed twist function $\phi_{\epsilon}(z)$ for
  $\epsilon \in [0, 1]$.}
\label{fig: poles}
\end{figure}

The twist function $\phi_{g{\rm FR}}(z) = 1$ was shown in \cite{Delduc:2012vq}
to correspond to a certain generalisation for the superstring of
the Faddeev-Reshetikhin Poisson bracket \cite{Faddeev:1985qu}. To deform the
superstring $\sigma$-model we will  use the $\R$-matrix
\eqref{R twist gen} with the twist function $\phi = \phi_{\epsilon}$, depending on a real parameter $\epsilon$, defined by
\begin{equation} \label{def twist inv}
\phi_{\epsilon}(z)^{-1} = \phi_{\rm string}(z)^{-1} + \epsilon^2 \phi_{g{\rm FR}}(z)^{-1}.
\end{equation}
We shall also denote the corresponding Poisson bracket as $\{ \cdot, \cdot \}_{\epsilon}$.
The undeformed case is recovered in the limit $\epsilon \to 0$.
Recall from \eqref{extract gX} that in this limit the poles of
the twist function $\phi_{\rm string}$ play an important role
in extracting both the group valued field and the non-local
conserved charges of the superstring $\sigma$-model. We
shall extend this key observation to the deformed case
in order to extract the group valued field $g$ and the
non-local charges of the deformed theory from the poles
of the deformed twist function $\phi_{\epsilon}$. A natural
parametrisation for these poles is obtained by introducing  
 $\theta \in [-\frac{\pi}{4}, \frac{\pi}{4}]$ as
\begin{equation} \label{epsilon-theta}
\epsilon =   \sin(2 \theta).
\end{equation}
Note that with the chosen range of values of $\theta$, the
original deformation parameter lies in the range $\epsilon \in [-1, 1]$.
The reason for this apparent restriction is that, as we shall see, the points  
$\epsilon = \pm 1$ will play a special role in the deformation.
In terms of the new parametrisation \eqref{epsilon-theta}, the deformed twist function defined by \eqref{def twist inv} explicitly reads
\begin{equation} \label{def twist}
\phi_{\epsilon}(z) = \frac{4 z^4}{\prod_{k=0}^3 (z - i^k e^{i \theta})(z - i^k e^{-i \theta})}.
\end{equation}
Therefore the poles of this twist function lie at $e^{\pm i \theta}$ and their
images under multiplication by $i$, as depicted in figure \ref{fig: poles}.

Before proceeding to extract the fields $g, X$ from the behaviour
of the Lax matrix at these points, let us comment on the possibility of
further deforming the twist function \eqref{def twist}.
One could try to introduce a second real deformation parameter $r$ by
considering the twist function
\begin{equation} \label{def twist 2}
\phi_{r, \theta}(z) = \frac{4 z^4}{\prod_{k=0}^3 (z - i^k r e^{i \theta})(z - i^k r e^{-i \theta})}.
\end{equation}
In fact, this is the most general real deformation of $\phi_{\epsilon}(z)$, since we must require that the set of eight simple poles of the twist function be invariant under multiplication by $i$ as well as under complex conjugation.
Let us denote the corresponding Poisson bracket, defined in
the same way as \eqref{Lax PB}, by $\{ \cdot, \cdot \}_{r, \theta}$. It is
natural to ask what linear combination of Poisson brackets gives rise to
it. For this we simply need to invert   \eqref{def twist 2} which yields
\begin{equation*}
\phi_{r, \theta}(z)^{-1} = \phi_{\rm string}(z)^{-1} + \ha \big(1 - r^4 \cos(4 \theta) \big)
 \phi_{g{\rm FR}}(z)^{-1} +\qa (r^8 - 1) z^{-4}.
\end{equation*}
The twist function \eqref{def twist 2} is thus
a double deformation of $\phi_{\rm string}(z)^{-1}$ by
$\phi_{g{\rm FR}}(z)^{-1} = 1$ and $z^{-4}$. However, note that
we have the relation $\phi_{r, \theta}(z) = r^{-4} \phi_{\epsilon}(z / r)$. But a
rescaling of the spectral parameter such as $z \mapsto z / r$ simply
corresponds to a linear redefinition of the fields $A^{(i)}$ and $\Pi^{(i)}$.
Therefore the Poisson bracket $\{ \cdot, \cdot \}_{r, \theta}$ can be
obtained from the Poisson bracket $\{ \cdot, \cdot \}_{\epsilon}$,
associated with the twist \eqref{def twist inv}, by this linear
redefinition of the fields and an overall rescaling by $r^4$.
We thus conclude that any further deformation of the twist function \eqref{def twist} will not lead to a more general deformed model.

\subsection{The Hamiltonian}

Recall that we wish to keep the dynamics of the fields
   $A$ and $\Pi$ intact so as to preserve integrability.
In other words we want the dynamics of the Lax matrix
\eqref{Lax matrix} to still take the form of the zero curvature
equation \eqref{ZC eq}. We actually find that for any functional $f$ of the
phase space variables  $A$ and $\Pi$,
\begin{equation*}
\{ H_{\rm string}, f \}_{\epsilon} \simeq \{ H_{\rm string}, f \},
\end{equation*}
when taking into account the constraints \eqref{constraints}. In other words,
this equality holds up to terms proportional to the constraints.

\subsection{The group valued field}

In the superstring $\sigma$-model, the fields $g$ and
$X$ can be obtained from the behaviour of the Lax
matrix at $1$ by means of the relation \eqref{extract gX}.
The significance of the point $z = 1$ is that it corresponds
to a double pole of the twist function $\phi_{\rm string}(z)$.
Having introduced a deformed twist function $\phi_{\epsilon}(z)$,
our next goal is to extract new fields $g$ and $X$ in a similar
fashion to \eqref{extract gX} but from the poles of $\phi_{\epsilon}(z)$.
However, as the deformation is turned on, the double pole at
$z = 1$ splits into two single poles at $z = e^{i \theta}$ and
$z = e^{- i \theta}$. We should therefore consider the behaviour
  of the Lax matrix at both these points (see figure \ref{fig: poles}).

\paragraph{Definition of $g$.} We define the group valued field
$g$ of the deformed theory by generalising the approach in
\cite{Delduc:2013fga} to the case at hand.
Since we want to describe a deformation of
the group valued field of the superstring $\sigma$-model, which takes
values in $F = \exp \f$, it is natural to require our field $g$ also to live in
$F$ for any value of the deformation parameter $\epsilon$. We
define $\partial_{\sigma} g g^{-1}$ to be the component along $\f$ relative
to the decomposition \eqref{decomposition} of $- g \L(e^{i \theta}) g^{-1}$.
In other words, we define $g \in F$ as the parameter of a gauge
transformation such that the gauge transformed Lax matrix
\begin{equation*}
\L^g(e^{i \theta}) = \partial_{\sigma} g g^{-1} + g \L(e^{i \theta}) g^{-1}
\end{equation*}
belongs to $\h_0 \oplus \n \subset \b$.
Now the fields $A^{(i)}$ and $\Pi^{(i)}$ of the model take values
in $\f$ which means that $A^{(i)} = \tau(A^{(i)})$, $\Pi^{(i)} = \tau(\Pi^{(i)})$
  with $\tau$ defined by  \eqref{psi reality}.
By virtue of these reality conditions and the definition
\eqref{Lax matrix} of the Lax matrix in terms of these fields we obtain
\begin{equation} \label{reality Lax 1}
\tau\big( \L(z) \big) = \L(\bar{z}).
\end{equation}
This, in particular, implies $\L^g(e^{-i \theta}) =
\tau\big(\L^g(e^{i \theta})\big)$ so that $\L^g(e^{-i \theta})$
 belongs to $\h_0 \oplus \tau(\n) \subset \tau(\b)$. Thus the
field $g$ is characterised by the single property
\begin{equation} \label{g def}
\L^g(e^{i \theta}) \in \h_0 \oplus \n.
\end{equation}
A nice feature of this definition is that in the limit
$\epsilon \to 0$, or equivalently $\theta \to 0$,
where the points $e^{i \theta}$ and $e^{-i \theta}$
both tend to $1$, we recover the defining relation
$\L^g(1) = 0$ of the $F$-valued field $g$ of the
superstring $\sigma$-model.
Indeed, in this limit we find that $\L^g(1) = \tau\big(\L^g(1)\big)$
from which it follows that $\L^g(1) \in \b \cap \tau(\b) = \h$.  In fact
$\L^g(1) \in \h_0 \subset \h$ which means that $\L^g(1) = -
 \tau\big(\L^g(1)\big)$. The only possibility is therefore that $\L^g(1) = 0$.

\paragraph{Definition of $X$.} We may also define the $\f$-valued field $X$ of the deformed theory by generalising the analysis of \cite{Delduc:2013fga}. Specifically, we set
\begin{equation} \label{X def}
X = \frac{i}{2 \gamma} \big( \L^g(e^{i \theta}) - \tau\big( \L^g(e^{i \theta}) \big) \big),
\end{equation}
where the real normalisation constant $\gamma$ will be fixed later. The field $X$ then takes
values in $\f$ because $\gamma$ is taken to be real and $\tau$ is an anti-linear involution, which implies that $\tau(X) = X$.
  We will come back to the limit $\epsilon \to 0$ of \eqref{X def} after fixing the value of $\gamma$ as a function of $\epsilon$.

Applying the linear operator \eqref{R def} to \eqref{X def} we find
\begin{equation} \label{RX eq}
RX = \frac{1}{2 \gamma} \big( \L^g(e^{i \theta}) + \tau\big( \L^g(e^{i \theta}) \big) \big).
\end{equation}
Combining \eqref{X def} with \eqref{RX eq} and using the fact that $\L^g(e^{- i \theta}) = \tau\big( \L^g(e^{i \theta}) \big)$ we therefore arrive at
\begin{equation} \label{Lg Rpmi}
\L^g(e^{\pm i \theta}) = \gamma (R \mp i) X.
\end{equation}

\subsection{Lifting to $(g,X)$}

Recall that in the superstring $\sigma$-model, the field $g$
describing the embedding of the string in target space and the field $X$ are related to the fields $A$ and
$\Pi$ entering the definition of the Lax matrix as
\begin{equation} \label{APi gX string}
A = - g^{-1} \partial_{\sigma} g, \qquad
\Pi = - g^{-1} X g.
\end{equation}
Further projecting these relations onto the various graded components of $\f$ yields equations for the fields $A^{(i)}$ and $\Pi^{(i)}$.
As already emphasised at the beginning of this section, in order to
 ensure that integrability is preserved throughout the deformation,
we have deformed neither the Lax matrix nor the  actual dynamics
of the fields $A^{(i)}$ and $\Pi^{(i)}$.
Now that we have candidates for the deformation of the
fields $g$ and $X$, what we need is to relate them to the
fields $A^{(i)}$ and $\Pi^{(i)}$. This will, in particular, enable us to
obtain the dynamics of  $g$ in the deformed theory. In other words, we
seek a deformation of the relations \eqref{APi gX string}. This can
be once more extracted from the behaviour of the Lax matrix at the
pair of points $e^{\pm i \theta}$.

Using   the relation \eqref{Lg Rpmi}, we can express the Lax matrix at $e^{\pm i \theta}$ as follows
\begin{equation} \label{APi pre}
\L(e^{\pm i \theta}) = - g^{-1} \partial_{\sigma} g + \gamma \, g^{-1} \big( (R \mp i) X \big) g.
\end{equation}
On the other hand, the left hand side can be evaluated directly in terms of the fields $A^{(i)}$ and $\Pi^{(i)}$ from the definition \eqref{Lax matrix} of the Lax matrix. Therefore \eqref{APi pre} constitutes a set of two equations relating $(g, X)$ to $(A, \Pi)$, each of which can be projected onto the four different gradings of $\f$. This yields a linear system of eight equations in the eight unknowns $A^{(i)}$, $\Pi^{(i)}$ for $i = 0, \ldots, 3$. Solving this system we finally arrive at the desired deformation of equations \eqref{APi gX string}, namely  
\begin{subequations} \label{APi to gX}
\begin{align}
A^{(0)} &= P_0 \biggl( - g^{-1} \partial_{\sigma} g + \gamma \, g^{-1} \biggl( \biggl(R + \frac{2 \eta}{1 - \eta^2} \biggr) X \biggr) g \biggr),\\
A^{(1)} &= \frac{\sqrt{1 + \eta^2}}{1 - \eta^2} P_1 \big( - g^{-1} \partial_{\sigma} g + \gamma \, g^{-1} \big( (R - \eta) X \big) g \big),\\
A^{(2)} &= \frac{1 + \eta^2}{1 - \eta^2} P_2 \big( - g^{-1} \partial_{\sigma} g + \gamma \, g^{-1} (R X) g \big),\\
A^{(3)} &= \frac{\sqrt{1 + \eta^2}}{1 - \eta^2} P_3 \big( - g^{-1} \partial_{\sigma} g + \gamma \, g^{-1} \big( (R + \eta) X \big) g \big),\\
\Pi^{(0)} &= - \gamma \frac{(1 + \eta^2)^2}{2 \eta (1 - \eta^2)} \, P_0 ( g^{-1} X g ),\\
\Pi^{(1)} &= \eta^2 \frac{\sqrt{1 + \eta^2}}{2 (1 - \eta^2)} P_1 \big( - g^{-1} \partial_{\sigma} g + \gamma \, g^{-1} \big( (R - \eta^{-3}) X \big) g \big),\\
\Pi^{(2)} &= - \gamma \frac{1 + \eta^2}{2 \eta} \, P_2( g^{-1} X g ),\\
\Pi^{(3)} &= - \eta^2 \frac{\sqrt{1 + \eta^2}}{2 (1 - \eta^2)} P_3 \big( - g^{-1} \partial_{\sigma} g + \gamma \, g^{-1} \big( (R + \eta^{-3}) X \big) g \big).
\end{align}
\end{subequations}
In these expressions, $P_i$ denote the projectors onto the subspaces $\f^{(i)}$ of $\f$.
Here we have introduced a new parameter $\eta$ related to the deformation
parameter $\epsilon$ as  
\begin{equation} \label{eta def}
\eta = - \frac{\epsilon}{1 + \sqrt{1 - \epsilon^2}}.
\end{equation}
Recall that the variable $\gamma$ was introduced in \eqref{X def} as
an overall factor in the definition of $X$. Remarkably, it turns out that
if we choose it to depend on the deformation parameter as follows  
\begin{equation} \label{gamma def}
\gamma = - \epsilon \sqrt{1 - \epsilon^2} = 2 \eta \frac{(1-\eta^2)}{(1+\eta^2)^2},
\end{equation}
then the  deformed Poisson brackets between the fields
$A^{(i)}$ and $\Pi^{(i)}$ follow from the canonical Poisson brackets
between $g$ and $X$ identical to those in \eqref{PB gX}.
Furthermore, with the dependence of $\gamma$ now fixed by \eqref{gamma def}, we can proceed to determine the limit $\epsilon \to 0$ of the definition \eqref{X def}. And indeed we find that it correctly reduces to the definition in the original superstring $\sigma$-model, namely the second relation in \eqref{extract gX}.

\subsection{Behaviour at $\epsilon = \pm 1$} \label{ds5h5}

To close this section, we consider the deformed model for the values  
$\epsilon = \pm 1$. The situation here is similar to the one discussed in the
 bosonic case \cite{Delduc:2013fga}. Specifically, we find that these values
  of $\epsilon$ correspond to an undeformed semi-symmetric space
$\sigma$-model. A first indication of this behaviour can be seen from
figure \ref{fig: poles}: for $\epsilon= \pm 1$, the deformed twist function once
again acquires four double poles at $z=e^{\pm i \frac{\pi}{4}}$ and
$z=e^{\pm  3 i \frac{\pi}{4}}$. Furthermore, in the neighbourhood of the
pole $z= e^{i \frac{\pi}{4}}$, the Lax matrix \eqref{Lax matrix} has
the behaviour
\begin{multline}
\L(z) = A^{(0)} + \ha e^{i \frac{\pi}{4}} A^{(1)} - \ha e^{3 i \frac{\pi}{4}} A^{(3)} + \Pi^{(0)} - e^{i \frac{\pi}{4}} \Pi^{(1)} - i \Pi^{(2)} - e^{3 i \frac{\pi}{4}} \Pi^{(3)}\\
+ \big( z - e^{i \frac{\pi}{4}} \big) \big( \mbox{\small $\frac{3}{2}$} A^{(1)}
+ 2 e^{i \frac{\pi}{4}} A^{(2)} +
 \mbox{\small $\frac{3}{2}$} i A^{(3)} - 2 e^{3 i \frac{\pi}{4}} \Pi^{(0)}
+ \Pi^{(1)} - i \Pi^{(3)} \big) + O\big( (z - e^{i \frac{\pi}{4}})^2 \big). \label{Latpisur4}
\end{multline}
In order to compare this situation to the undeformed one at $\epsilon=0$,
we introduce
$\hat{z} = e^{-i \frac{\pi}{4}} z$. The pole $z= e^{i \frac{\pi}{4}}$ then corresponds
to $\hat{z}=1$ and the expression \eqref{Latpisur4} takes the same form as in equation \eqref{Latzegal1},
\beqz
\hat{A} - 2 (\hat{z}-1) \hat{\Pi} + O((\hat{z}-1)^2)
\eeqz
provided we define
\begin{subequations} \label{ahpih}
\begin{align}
\hat{A} &= A^{(0)}  + \Pi^{(0)} +  e^{i \frac{\pi}{4}}  ( \ha A^{(1)} -  \Pi^{(1)} )
- i \Pi^{(2)}  + e^{-i \frac{\pi}{4}} ( \ha A^{(3)} +  \Pi^{(3)} ),\\
\hat{\Pi} &= -\Pi^{(0)}  - \ha e^{i \frac{\pi}{4}}  (\mbox{\small $\frac{3}{2}$}
A^{(1)} +   \Pi^{(1)} ) - i A^{(2)} + \ha e^{-i \frac{\pi}{4}}
 (\mbox{\small $\frac{3}{2}$}   A^{(3)} -   \Pi^{(3)} ).
\end{align}
\end{subequations}
One can also check that the Poisson algebra satisfied by the fields
$\hat{A}$ and $\hat{\Pi}$ corresponds to the undeformed one,
given in \eqref{PB string}. Furthermore, the constraints
\eqref{constraints}  take the same form when expressed
in terms of $(\hat{A},\hat{\Pi})$, namely
\begin{align*}
\C^{(0)} &= \Pi^{(0)}= -  \hat{\Pi}^{(0)} ,\\
\C^{(1)} &= \ha A^{(1)} + \Pi^{(1)} = -  e^{-i \frac{\pi}{4}} (\ha \hat{A}^{(1)} + \hat{\Pi}^{(1)} ),\\
\C^{(3)}&= -\ha A^{(3)} + \Pi^{(3)} = -  e^{i \frac{\pi}{4}} (-\ha \hat{A}^{(3)}
+ \hat{\Pi}^{(3)}),\\
T_\pm &= \str(A_\pm^{(2)} A_\pm^{(2)}) = -  \str(\hat{A}_\pm^{(2)} \hat{A}_\pm^{(2)}).
\end{align*}
However, the fields $\hat{A}$ and $\hat{\Pi}$ satisfy different reality conditions from the fields $A$ and $\Pi$. Recall that the latter belong to $\mathfrak{f} = (\Gr \otimes \mathfrak{su}(2,2|4))^{[0]}$, where an element $M$ in $\mathfrak{su}(2,2|4)$ satisfies the reality condition $\tau(M)=M $ with $\tau$ the antilinear map defined by \eqref{psi reality}.
Starting from an element $M$ in $\mathfrak{su}(2,2|4)$, formulas \eqref{ahpih} suggest to consider the following element of $\mathfrak{sl}(4\vert 4)$,
\begin{equation} \label{hat M}
\hat M=M^{(0)}+e^{i\frac{\pi}{4}}M^{(1)}+e^{i\frac{\pi}{2}}M^{(2)}+
e^{i\frac{3\pi}{4}}M^{(3)}.
\end{equation}
Using the reality conditions for  $M$ and the anti-linearity of $\tau$, one
finds that
\begin{align*}
\tau(\hat M) &=M^{(0)}+e^{-i\frac{\pi}{4}}M^{(1)}+e^{-i\frac{\pi}{2}}M^{(2)}+e^{-i\frac{3\pi}{4}}M^{(3)},\\
& ={\hat M}^{(0)}+e^{-i\frac{\pi}{2}}{\hat M}^{(1)}+e^{-i\pi}{\hat M}^{(2)}+e^{-i\frac{3\pi}{2}}{\hat M}^{(3)} = \Omega^{-1}(\hat M).
\end{align*}
The last equality is obtained by using the property \eqref{propOmega} of the
 automorphism  $\Omega$ defining the $\mathbb{Z}_4$-grading of
$\mathfrak{su}(2,2|4)$.
Thus, the reality condition for the element \eqref{hat M} may be written as
\begin{equation}
\Omega \circ \tau(\hat M)={\hat M}. \label{eqnewreality}
\end{equation}
One can check that $(\Omega \circ \tau)^2$ is equal to the identity.
Working in the fundamental representation of $\mathfrak{sl}(4|4)$, the
reality condition \eqref{eqnewreality} may be written more explicitly as
\begin{equation} \label{staralg}
 {\hat{\mathbf K}}^{-1}{\hat M}^*{\hat{\mathbf K}}={\hat M},
\qquad {\hat{\mathbf K}} = \text{diag}( \mathsf{k}, -\mathsf{k}, -i\mathsf{k}, i\mathsf{k}),
\end{equation}
where $\mathsf{k}$ is defined in \eqref{auto Omega}.
The matrix ${\hat{\mathbf K}}$ is antisymmetric and non-singular.
Up to conjugation, equation ({\ref{staralg}) means that the matrix
$\hat M$ belongs to the real superalgebra $\mathfrak{su}^{\ast}(4|4)$
as defined in \cite{Serganova:1983vp}. In particular, the fields $\hat A$
and $\hat\Pi$ belong to the real form
$(\Gr \otimes \mathfrak{su}^{\ast}(4|4))^{[0]}$ of the Lie
algebra $\f^{\mathbb{C}}$. The bosonic subalgebra of
$\mathfrak{su}^{\ast}(4|4)$, after projection, is
$\mathfrak{su}^{\ast}(4)\oplus \mathfrak{su}^{\ast}(4)\equiv
\mathfrak{so}(1,5)\oplus \mathfrak{so}(1,5)$. Notice that the
grade zero part, and thus the gauge algebra, is not modified.

We thus find that at $\epsilon = \pm 1$, we obtain again an undeformed $\sigma$-model
on the semi-symmetric space $PSU^\ast(4|4)/(SO(4,1) \times SO(5))$ with
bosonic sector corresponding to $dS_5 \times H^5$ as announced in
\cite{Delduc:2013qra}.

\section{$q$-deformed symmetry algebra}
\label{sec: qdefsymmetry}
Recall from section \ref{sec: sym undef} that charges generating the
$F$ symmetry of the undeformed superstring $\sigma$-model on $F/G$
can be extracted from the expansion of the gauge transformed monodromy
matrix at the pole $z = 1$ of the undeformed twist function. Since the
monodromy matrix is still conserved in the deformed theory   by virtue of
the zero curvature equation \eqref{ZC eq} not being modified, it makes sense
to try to extract the global charges of the deformed model in a similar way.

Specifically, we consider the gauge transformed monodromy matrix $T^g(z)$
at the poles $z = e^{\pm i \theta}$ of the deformed twist function
\begin{equation} \label{Tg expr}
T^g\big( e^{\pm i \theta} \big) = g(\infty) T\big( e^{\pm i \theta} \big) g(- \infty)^{-1}
= P \overleftarrow{\exp} \left[ \int_{- \infty}^{\infty} d\sigma \L^g\big( e^{\pm i \theta} \big) \right].
\end{equation}
We note here the first difference with the behaviour of $T^g(z)$ near $z = 1$
in \eqref{Tg near 1}: since the gauge transformed Lax matrix $\L^g(z)$ does
not vanish at $z = e^{\pm i \theta}$,   by \eqref{Lg Rpmi}, the expansion of
$T^g(z)$ near these points is already non-trivial at leading order.

In order to evaluate the right hand side of \eqref{Tg expr} further we recall
from \eqref{Lg Rpmi} and \eqref{R pm i} that the expressions $\L^g(e^{i \theta})$
and $\L^g(e^{- i \theta})$ respectively take values in the subalgebras
$\h_0 \oplus \n$ and $\h_0 \oplus \tau(\n)$. We may therefore write them as follows
\begin{equation*}
\L^g\big( e^{i \theta} \big) = \gamma \sum_{\mu = 1}^7 h_{\mu} H^{\mu} + \gamma \sum_{\alpha > 0} e_{\alpha} E^{\alpha}, \qquad
\L^g\big( e^{- i \theta} \big) = - \gamma \sum_{\mu = 1}^7 \tilde{h}_{\mu} H^{\mu} - \gamma \sum_{\alpha > 0} e_{-\alpha} E^{-\alpha},
\end{equation*}
for some $\Gr$-valued fields $h_{\mu}(\sigma)$, $\tilde{h}_{\mu}(\sigma)$
and $\Gr^{\mathbb{C}}$-valued fields $e_{\pm \alpha}(\sigma)$ such that
 $|h_{\mu}(\sigma)| = |\tilde{h}_{\mu}(\sigma)| = 0$ and $|e_{\pm \alpha}(\sigma)| = |E^{\alpha}|$. By using the reality condition $\L^g(e^{-i \theta}) = \tau\big( \L^g(e^{i \theta}) \big)$ we then find that these fields are related as (see appendix \ref{app: psu})
\begin{equation} \label{h e reality}
\tilde{h}_{\mu} = h_{\mu}^{\ast} = h_{\mu}, \qquad
e_{\epsilon_a - \epsilon_b}^{\ast} = (-1)^{s(a)+s(b)} i^{|E_{ab}|} e_{\epsilon_b - \epsilon_a},
\end{equation}
for $1 \leq \mu \leq 7$ and all positive roots $\epsilon_a - \epsilon_b \in \Phi^+$.
Following the same reasoning as in the bosonic case
 \cite{Delduc:2013fga}, the Cartan direction in \eqref{Tg expr} may
be factored out as
\begin{subequations} \label{Tg deformed}
\begin{align}
\label{Tg deformed +} T^g\big( e^{i \theta} \big) &= \exp \left( \gamma \int_{-\infty}^{\infty} d\sigma \sum_{\mu=1}^7 h_{\mu}(\sigma) H^{\mu} \right) P \overleftarrow{\exp} \left[ \gamma \sum_{\alpha > 0} \int_{- \infty}^{\infty} d\sigma \, \mathfrak{J}^E_{\alpha}(\sigma) E^{\alpha} \right],\\
\label{Tg deformed -} T^g\big( e^{-i \theta} \big) &= P \overleftarrow{\exp} \left[ - \gamma \sum_{\alpha > 0} \int_{- \infty}^{\infty} d\sigma \, \mathfrak{J}^E_{-\alpha}(\sigma) E^{- \alpha} \right] \exp \left( - \gamma \int_{-\infty}^{\infty} d\sigma \sum_{\mu=1}^7 h_{\mu}(\sigma) H^{\mu} \right).
\end{align}
\end{subequations}
The notation in these expressions is as follows. For any positive
root $\alpha > 0$ we define the fields
\begin{equation} \label{JH JE}
\mathfrak{J}^H_{\alpha}(\sigma) = \sum_{\mu = 1}^7 \alpha(H^{\mu}) h_{\mu}(\sigma), \qquad
\mathfrak{J}^E_{\pm \alpha}(\sigma) = e_{\pm \alpha}(\sigma) e^{- \gamma \chi_{\alpha}(\sigma)} e^{\gamma \chi_{\alpha}(\mp \infty)}.
\end{equation}
Moreover, the function $\chi_{\alpha}$ for $\alpha > 0$ is explicitly defined as
\begin{equation} \label{chi def}
\chi_{\alpha}(\sigma) = \ha \int^{\infty}_{- \infty} d\sigma' \epsilon_{\sigma \sigma'} \mathfrak{J}^H_{\alpha}(\sigma'),
\end{equation}
where $\epsilon_{\sigma \sigma'} = \text{sgn}(\sigma - \sigma')$, $\text{sgn}$
being the sign function,
  which satisfies $\partial_{\sigma} \epsilon_{\sigma \sigma'} = 2 \delta_{\sigma \sigma'}$. By construction this satisfies $\partial_{\sigma} \chi_{\alpha}(\sigma) = \mathfrak{J}^H_{\alpha}(\sigma)$ and takes the following values at infinity
\begin{equation} \label{chi inf}
\chi_{\alpha}(\pm \infty) = \pm \ha \int_{-\infty}^{\infty} d\sigma' \mathfrak{J}^H_{\alpha}(\sigma').
\end{equation}
As in the case of bosonic $\sigma$-models \cite{Delduc:2013fga},
  it can be deduced from the conservation of $T^g(z)$ and its explicit value \eqref{Tg deformed} at the points $z = e^{\pm i \theta}$ that the charges
\begin{equation} \label{cons charges}
\int_{-\infty}^{\infty} d\sigma \, \mathfrak{J}^H_{\alpha_{\mu}}(\sigma), \qquad
\int_{-\infty}^{\infty} d\sigma \, \mathfrak{J}^E_{\pm \alpha_{\mu}}(\sigma)
\end{equation}
are separately conserved for each simple root $\alpha_i$. Note also that the conservation of the former would also follow from the conservation of $\int_{-\infty}^{\infty} d\sigma \, h_{\mu}(\sigma)$ using the first relation in \eqref{JH JE} between the densities $\mathfrak{J}^H_{\alpha_{\mu}}(\sigma)$ and $h_{\mu}(\sigma)$.

\paragraph{Deformed symmetry algebra.} We now wish to derive the
Poisson algebra of the charges \eqref{cons charges}. As in the undeformed
case, this can be obtained from the Poisson bracket \eqref{PB gX c} of
the field $X$ with itself since  the charge densities \eqref{JH JE} are
entirely defined in terms of the components of the field $X$.
Indeed, by definition \eqref{X def} of the field $X$, it may be written more
explicitly as
\beq
X = \sum_{\mu = 1}^7 i h_{\mu} H^{\mu} + \frac{i}{2} \sum_{\alpha > 0}
\big( e_{\alpha} E^{\alpha} + e_{-\alpha} E^{-\alpha} \big). \label{another def X}
\eeq
Using the expression \eqref{psl Casimir} for the tensor Casimir we
may write the Poisson bracket \eqref{PB gX c} as
\begin{align*}
\big\{ X_{\1}(\sigma), X_{\2}(\sigma') \big\}_{\epsilon} &= \bigg( \sum_{\mu, \nu = 1}^7
Y_{\mu\nu} H^{\mu} \otimes [ H^{\nu}, X(\sigma)]\\
&\qquad\qquad\qquad + \sum_{\alpha > 0} \big( (-1)^{|E^{\alpha}|} E^{\alpha} \otimes [ E^{-\alpha}, X(\sigma) ] + E^{-\alpha} \otimes [ E^{\alpha}, X(\sigma)] \big) \bigg) \delta_{\sigma \sigma'}.
\end{align*}
By comparing coefficients of the various basis elements of
$\mathfrak{sl}(4|4)$ in the first tensor factor on both sides we find
\begin{subequations} \label{PB hX eX}
\begin{align}
\label{PB hX} i \{ h_{\mu}(\sigma), X(\sigma') \}_{\epsilon} &= \sum_{\nu = 1}^7 Y_{\mu\nu} [ H^{\nu}, X(\sigma) ] \delta_{\sigma \sigma'}, \\
\label{PB eX} i \{ e_{\alpha}(\sigma), X(\sigma') \}_{\epsilon} &= 2 \, [ E^{-\alpha}, X(\sigma) ] \delta_{\sigma \sigma'}.
\end{align}
\end{subequations}
Multiplying the first of these equations by the symmetrised Cartan matrix and using the relation \eqref{BX prod} along with the fact that $I = \sum_{\nu = 1}^7 x_{\nu} H^{\nu}$ yields
\begin{equation*}
i \sum_{\rho = 1}^7 B_{\mu \rho} \{ h_{\rho}(\sigma), X(\sigma') \}_{\epsilon} = [ H^{\mu}, X(\sigma) ] \delta_{\sigma \sigma'} - \omega^{-1} \alpha_{\mu}(H^8) [ I, X(\sigma) ] \delta_{\sigma \sigma'}.
\end{equation*}
However, since the generator $I$ is central in $\mathfrak{sl}(4|4)$, the second term on the right hand side vanishes. Using the definition \eqref{JH JE} we are left with
\begin{equation*}
i \{ \mathfrak{J}^H_{\alpha_{\mu}}(\sigma), X(\sigma') \}_{\epsilon} = [ H^{\mu}, X(\sigma) ] \delta_{\sigma \sigma'}.
\end{equation*}
Consider the component of this equation along the Cartan subalgebra. Since the right hand side involves only non-Cartan generators, it follows that
\begin{equation} \label{jh jh pb}
i \{ \mathfrak{J}^H_{\alpha_{\mu}}(\sigma), \mathfrak{J}^H_{\alpha_{\nu}}(\sigma') \}_{\epsilon} = 0,
\end{equation}
for any $1 \leq \mu, \nu \leq 7$.
Likewise, by comparing the coefficient of $E^{\pm \alpha_{\nu}}$ on both sides of this equation we obtain
\begin{equation} \label{jHepbe}
i \{ \mathfrak{J}^H_{\alpha_{\mu}}(\sigma), e_{\pm \alpha_{\nu}}(\sigma') \}_{\epsilon} = \pm B_{\mu \nu} \, e_{\pm \alpha_{\nu}}(\sigma) \delta_{\sigma \sigma'}.
\end{equation}
Using the definition \eqref{chi def} of $\chi_{\alpha}(\sigma)$, this in particular implies the following
\begin{equation} \label{chi e pb}
i \{ e^{- \gamma \chi_{\alpha_{\mu}}(\sigma)}, e_{\pm \alpha_{\nu}}(\sigma') \}_{\epsilon} = \mp \ha \gamma B_{\mu \nu} \, e_{\pm \alpha_{\nu}}(\sigma') e^{- \gamma \chi_{\alpha_{\mu}}(\sigma)} \epsilon_{\sigma \sigma'}.
\end{equation}
Furthermore, specialising \eqref{PB eX} to the case of a simple root $\alpha = \alpha_{\mu}$, and comparing the coefficient of $E^{- \alpha_{\nu}}$ on both sides we have
\begin{equation*}
i \{ e_{\alpha_{\mu}}(\sigma), e_{- \alpha_{\nu}}(\sigma') \}_{\epsilon} = 4 \, \partial_{\sigma} \chi_{\alpha_{\mu}}(\sigma) \delta_{\mu \nu} \delta_{\sigma \sigma'}.
\end{equation*}
Putting the above together we find that the Poisson brackets between the charge densities $\mathfrak{J}^H_{\alpha_{\mu}}(\sigma)$ and $\mathfrak{J}^E_{\pm \alpha_{\mu}}(\sigma)$ take the form
\begin{subequations} \label{JHJE PB}
\begin{align}
i \{ \mathfrak{J}^E_{\alpha_{\mu}}(\sigma), \mathfrak{J}^E_{- \alpha_{\nu}}(\sigma') \}_{\epsilon} &= - 2  \gamma^{-1} \partial_{\sigma} \big( e^{-2 \gamma \chi_{\alpha_{\mu}}(\sigma)} \big) \delta_{\mu \nu} \delta_{\sigma \sigma'},\\
i \{ \mathfrak{J}^H_{\alpha_{\mu}}(\sigma), \mathfrak{J}^E_{\pm \alpha_{\nu}}(\sigma') \}_{\epsilon} &= \pm B_{\mu \nu} \, \mathfrak{J}^E_{\pm \alpha_{\nu}}(\sigma) \delta_{\sigma \sigma'}.
\end{align}
\end{subequations}
We define the integrated charges as
\begin{equation} \label{QHE def}
Q^H_{\alpha_{\mu}} = \int_{-\infty}^{\infty} d\sigma  \mathfrak{J}^H_{\alpha_{\mu}}(\sigma), \qquad
Q^E_{\pm \alpha_{\mu}} = \left( \frac{\gamma}{4 \sinh \gamma} \right)^{\frac{1}{2}} \int_{-\infty}^{\infty} d\sigma \mathfrak{J}^E_{\pm \alpha_{\mu}}(\sigma),
\end{equation}
where the normalisation in $Q^E_{\pm \alpha_{\mu}}$ was introduced for convenience as in the bosonic case \cite{Delduc:2013fga}. With these definitions, the collection of Poisson brackets \eqref{jh jh pb} and \eqref{JHJE PB} for the densities now implies
\begin{subequations} \label{QHE rel}
\begin{align}
i \{ Q^H_{\alpha_{\mu}}, Q^H_{\alpha_{\nu}} \}_{\epsilon} &= 0,\\
i \{ Q^E_{\alpha_{\mu}}, Q^E_{- \alpha_{\nu}} \}_{\epsilon} &= \delta_{\mu \nu} \frac{q^{Q^H_{\alpha_{\mu}}} - q^{- Q^H_{\alpha_{\mu}}}}{q - q^{-1}},\\
i \{ Q^H_{\alpha_{\mu}}, Q^E_{\pm \alpha_{\nu}} \}_{\epsilon} &= \pm B_{\mu \nu} \, Q^E_{\pm \alpha_{\nu}},
\end{align}
\end{subequations}
where we have made use of the values \eqref{chi inf}. The new
deformation parameter $q$ used here is related to $\gamma$,
defined in \eqref{gamma def}, as follows  
\begin{equation} \label{q def}
q = e^{-\gamma} = \exp\left(  \epsilon \sqrt{1 - \epsilon^2} \right) =
\exp \left( - \frac{2 \eta (1 - \eta^2)}{(1 + \eta^2)^2} \right).
\end{equation}

\paragraph{Charges associated with non simple roots.}

In order to construct conserved charges $Q^E_{\alpha}$ associated with any positive root  $\alpha \in \Phi^+$, we make a choice of normal ordering on the set of positive roots $\Phi^+$ of $\mathfrak{psl}(4|4)$ (see for instance \cite{Khoroshkin_1991,Asherova_1979,Tolstoy_1989a}). The latter is defined as a partial ordering on $\Phi^+$ with the property that if $\alpha < \beta$ and $\alpha + \beta$ is a root, then $\alpha < \alpha + \beta < \beta$. Using such an ordering, the remaining path ordered   exponential  appearing on the right hand side of \eqref{Tg deformed +} can be expressed in terms of simple exponentials of individual generators $E^{\alpha}$. Specifically, we have
\begin{equation} \label{exp prod}
P \overleftarrow{\exp} \left[ \gamma \sum_{\alpha > 0} \int_{- \infty}^{\infty} d\sigma \, \mathfrak{J}^E_{\alpha}(\sigma) E^{\alpha} \right] = \sideset{}{^<}\prod_{\alpha > 0} \exp \left( \gamma \int_{-\infty}^{\infty} d\sigma \, \mathfrak{Q}^E_{\alpha}(\sigma) E^{\alpha} \right)
\end{equation}
where $\mathfrak{Q}^E_{\alpha}(\sigma)$ are $\Gr^{\mathbb{C}}$-valued fields
  whose parities are the ones of  $  E^{\alpha}$.
 In particular, given any simple root $\alpha_{\nu}$ we have $\mathfrak{Q}^E_{\alpha_{\nu}}(\sigma) = \mathfrak{J}^E_{\alpha_{\nu}}(\sigma)$. The ordering of the product on the right hand side of \eqref{exp prod} is determined by the normal ordering of the corresponding roots.
 Although the  latter is only a partial ordering on $\Phi^+$, there is no ambiguity in the above product since generators $E^{\alpha}$ and $E^{\beta}$ commute whenever the corresponding roots $\alpha$ and $\beta$ are not ordered.
Moreover, equation \eqref{exp prod} implies that for any pair of simple roots   $\alpha_{\mu} < \alpha_{\nu}$ such that $\alpha_{\mu} + \alpha_{\nu}$ is a root, we have
\beq \label{res21mas}
\mathfrak{Q}^E_{\alpha_{\nu} + \alpha_{\mu}}(\sigma) = \mathfrak{J}^E_{\alpha_{\nu} + \alpha_{\mu}}(\sigma) - \gamma \, N_{\alpha_{\mu}, \alpha_{\nu}} \mathfrak{J}^E_{\alpha_{\nu}}(\sigma) \int_{-\infty}^{\sigma} d\sigma' \mathfrak{J}^E_{\alpha_{\mu}}(\sigma'),
\eeq
with $N_{\alpha_\mu, \alpha_{\nu}}$ defined by \eqref{def of N}.
In this case, we define  the corresponding charge by
\begin{equation} \label{Q roots def}
Q^E_{\alpha_\nu +   \alpha_{\mu}} =   \frac{\gamma}{4 \sinh \gamma}
\int_{-\infty}^{\infty} d\sigma \, \mathfrak{Q}^E_{\alpha_\nu +  \alpha_{\mu}}(\sigma).
\end{equation}

\paragraph{$q$-Poisson-Serre relations.}

In this paragraph, we list the $q$-Poisson-Serre relations. They are proved in appendix \ref{app: qSerre}. To write these relations, we first define  the $q$-Poisson bracket of  any charges $Q^E_\alpha$ and $Q^E_\beta$ associated with positive roots $\alpha$ and $\beta$.
It is simply given by
\beq
\big( \text{ad}_{\{ \cdot, \cdot \}_{q \, \epsilon}} Q^E_{\alpha} \big)(Q^E_{\beta})
= \{ Q^E_{\alpha}, Q^E_{\beta} \}_{q \, \epsilon} =
\{ Q^E_{\alpha}, Q^E_{\beta} \}_{\epsilon} - i \gamma \, (\alpha, \beta) Q^E_{\alpha} Q^E_{\beta}. \label{XdefofqPS}
\eeq
The standard $q$-Poisson-Serre relations then take the form (see for
 instance \cite{Khoroshkin_1991}):
\begin{subequations} \label{all stand qPS}
\begin{align}
\label{all stand qPS a} \{ Q^E_{\alpha_\nu}, Q^E_{\alpha_\mu} \}_{\epsilon} &= 0 \qquad \mbox{when}
\qquad (\alpha_\nu,\alpha_\mu)=0,\\
\{ Q^E_{\alpha_\nu}, \{ Q^E_{\alpha_\nu} , Q^E_{\alpha_\mu} \}_{q \, \epsilon} \}_{q \, \epsilon} &= 0
\qquad \mbox{when} \qquad \alpha_\nu < \alpha_\mu \qquad \mbox{and} \qquad (\alpha_\nu,\alpha_\mu) \neq 0,\\
\{ \{ Q^E_{\alpha_\mu}, Q^E_{\alpha_\nu} \}_{q \, \epsilon}, Q^E_{\alpha_\nu} \}_{q \, \epsilon} &= 0
\qquad \mbox{when} \qquad \alpha_\nu > \alpha_\mu \qquad \mbox{and} \qquad (\alpha_\nu,\alpha_\mu) \neq 0.
\end{align}
\end{subequations}
In the first relation \eqref{all stand qPS a}, note that for such a pair  of simple roots,  the corresponding $q$-Poisson bracket \eqref{XdefofqPS}  is equal to the ordinary Poisson bracket. Let us also point out that this situation clearly holds when $\alpha_\nu =\alpha_\mu$ and $E^{\alpha_\mu}$ is odd.

These relations are written independently of the choice of Dynkin diagram.
We can however
be more precise, even without specialising to a particular  Dynkin diagram.
Indeed, for $\nu \neq \mu$, we only have $(\alpha_\nu,\alpha_\mu) \neq 0$
when $\nu = \mu \pm 1$.   The proof given in appendix \ref{app: qSerre}
 is for $\nu = \mu +1$ with the ordering  $\alpha_\mu < \alpha_\mu +
\alpha_{\mu +1} < \alpha_{\mu +1}$.  Let us note that we will also
prove the relation
\beq \label{qpsm21text}
\{ Q^E_{\alpha_{\mu} }, Q^E_{\alpha_{\mu+1}} \}_{q \, \epsilon} =
- 2 i N_{ \alpha_{\mu+1},\alpha_\mu} (\alpha_{\mu+1} , \alpha_\mu)
Q^E_{\alpha_{\mu}+\alpha_{\mu+1}}.
\eeq

For completeness, non-standard $q$-Poisson-Serre relations should
also be proved. For simplicity, this will be done in the case of the
standard Dynkin diagram. In this case, there is in fact just one
non-standard relation. It is associated with the part of the
Dynkin diagram shown on figure \ref{figure_root_odd}.
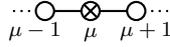
\begin{figure}[h]
\begin{center}
\begin{tikzpicture}[baseline =-5,scale=.6]
\draw[thick] (3,0) -- (5,0);
\draw[thick, dotted] (2.3,0) -- (3,0) node[below]{\scriptsize $\mu-1$ \;\;};
\draw[thick, dotted] (5,0) node[below]{\scriptsize \;\; $\mu+1$} -- (5.7,0);
\foreach \x in {3,4,5}
\filldraw[fill=white, thick] (\x,0) circle (2mm);
\draw[thick] (4,0)node[below=2]{\scriptsize $\mu$} ++(-.15,-.15)  -- ++(.3,.3);
\draw[thick] (4,0)++(.15,-.15) -- ++(-.3,.3);
\end{tikzpicture}
\end{center}
\caption{Part of the Dynkin diagram relevant for the non-standard
$q$-Poisson-Serre relation.}
\label{figure_root_odd}
\end{figure}
With these conventions, the relation is
\cite{Khoroshkin_1991,Floreanini:1991er,Dobrev:2009bb}
\beq \label{all non stand qPS}
\{ \{ Q^E_{\alpha_{\mu}}, Q^E_{\alpha_{\mu-1}} \}_{q \, \epsilon} ,
\{ Q^E_{\alpha_{\mu }}, Q^E_{\alpha_{\mu+1}} \}_{q \, \epsilon} \}_\epsilon =0.
\eeq
All these relations are proved in appendix \ref{app: qSerre}.

\medskip

To conclude this section, we have shown that the $\mathfrak{psu}(2,2|4)$ symmetry
of the $AdS_5 \times S^5$ superstring is replaced in the deformed theory by the
classical analog of the quantum group corresponding to this Lie superalgebra. 
This is a generalisation of the situation first encountered 
for the case of the squashed 3-sphere $\sigma$-model in 
\cite{Kawaguchi:2011pf,Kawaguchi:2012gp}. The relation \eqref{q def} 
between $q$ and $\eta$ is in agreement with the one found in 
\cite{Arutyunov:2013ega}. More precisely, in \cite{Arutyunov:2013ega}, 
$q=\exp(-\nu/g)$ with $\nu = 2 \eta/(1+\eta^2)$. The value of $g$ 
should then be fixed by comparing the 
 different prefactors in the Lagrangian (2.1) of \cite{Arutyunov:2013ega} 
and in the action \eqref{PRL Action} below. This comparison leads to
 \eqref{q def}. Finally, it is expected that the full symmetry algebra, including higher conserved charges, corresponds to the classical 
analog of $U_q(\widehat{\mathfrak{psu}(2,2|4)})$. This would again generalise 
the situation in the squashed sphere $\sigma$-model \cite{Kawaguchi:2012ve}.

\section{The deformed superstring action} \label{sec_action}

So far we have constructed a deformation
in the hamiltonian framework. We would now like to derive the corresponding action.
To do so, we need to perform the inverse Legendre
transform in the presence of constraints. The fields
$\lambda^+$ and $\lambda^-$ will be treated as spectator fields
in the inverse Legendre transform.
The starting point is   to consider the quantity  (see for instance
 \cite{henneauxteitelboim_1994})
\begin{multline} \label{eqtlvar}
\int d\sigma d\tau \Bigl[- \str\left(g^{-1} \partial_\tau g
g^{-1}X g\right) -\\
 \bigl(\lambda^+ \mathcal{T}_+ + \lambda^- \mathcal{T}_-
- \str(\mu^{(3)} \C^{(1)}) - \str(\mu^{(1)} \C^{(3)}) - \str\big( (A^{(0)} +
 \ell) \Pi^{(0)} \big) \Bigr]
\end{multline}
as a functional of $g$, $X$, $\mu^{(1)}$, $\mu^{(3)}$ and $\ell$.
Performing the Legendre transform then corresponds to extremizing
\eqref{eqtlvar} with respect to all the fields except $g$.
 Specifically, varying \eqref{eqtlvar} with respect to the Lagrange multipliers
$\ell$, $ \mu^{(3)}$ and
$\mu^{(1)}$     produces   the bosonic constraint $\C^{(0)}  = -(g^{-1} X g)^{(0)}
 \simeq 0$,
and the fermionic constraints $\C^{(1)} \simeq 0$ and $\C^{(3)}
\simeq 0$, where  
\begin{subequations} \label{C1 C3}
\begin{align}
{\cal C}^{(1)} &=-\frac{1}{\sqrt{1+\eta^2}}\left((g^{-1}Xg)^{(1)} -
\eta(g^{-1}RXg)^{(1)}+\frac{(1+\eta^2)^2}{2(1-\eta^2)}(g^{-1}
\partial_\sigma g)^{(1)}\right),\\
{\cal C}^{(3)} &=-\frac{1}{\sqrt{1+\eta^2}}\left((g^{-1}Xg)^{(3)} +
\eta(g^{-1}RXg)^{(3)}-\frac{(1+\eta^2)^2}{2(1-\eta^2)}(g^{-1}
\partial_\sigma g)^{(3)}\right).
\end{align}
\end{subequations}
Here we have used the equations
 \eqref{constraints}
and the results \eqref{APi to gX}.
The equation obtained by taking the variation
 with respect to $X$ is analysed in the next paragraph.
We then get the deformed action  $S_\epsilon[g]$ by plugging all
these equations in \eqref{eqtlvar}.

\subsection{Relating $X$ to $g^{-1} \partial_{\tau} g$}

Using all the above,
 and setting the constraints $\C^{(0)}$, $\C^{(1)}$ and $\C^{(3)}$ to zero, the non-zero grades $i = 1,2,3$
  of the equation obtained by extremizing \eqref{eqtlvar} with
respect to $X$ can be shown to take the form
\begin{multline} \label{g eom}
D_{ii} \gamma^{0 \alpha} (g^{-1} \partial_{\alpha} g)^{(i)} + R^{(i)}_g \big( D_{22} (g^{-1} \partial_{\sigma} g)^{(2)} \big)\\
= (g^{-1} X g)^{(i)} - R^{(i)}_g R^{(2)}_g (g^{-1} X g) + R^{(i)}_g (g^{-1} X g)^{(1)} - R^{(i)}_g (g^{-1} X g)^{(3)}\\
- \frac{D_{11} \gamma^{00}}{\sqrt{1+\eta^2}} \big( R_g^{(i)} \mu^{(1)} - R_g^{(i)}
\mu^{(3)} + \mu^{(i)} \big),
\end{multline}
 where for notational simplicity we have introduced $\mu^{(2)} = 0$.
The rest of the notation is defined as follows. We have introduced the diagonal matrix
\begin{equation*}
D = \frac{(1 + \eta^2)^2}{2 (1 - \eta^2)} \, \text{diag}\bigg( 1, \frac{2}{1 - \eta^2}, 1 \bigg),
\end{equation*}
and the following shorthands  
\begin{equation}
R^{(1)}_g = - \eta P_1 \circ R_g, \qquad
R^{(2)}_g = - \frac{2 \eta}{1 - \eta^2} P_2 \circ R_g, \qquad
R^{(3)}_g = - \eta P_3 \circ R_g,
\end{equation}
 where
\begin{equation*}
R_g = \text{Ad}\, g^{-1} \circ R \circ \text{Ad} \, g.
\end{equation*}
As recalled above, to perform the inverse Legendre transform,
the last ingredient we need is an expression relating
  $X$ to the temporal derivative
$g^{-1} \partial_{\tau} g$ of the group valued field $g$. This can be
extracted from the field equations \eqref{g eom} along with the
constraints $\C^{(1)}$ and $\C^{(3)}$ as we now explain.
 Consider the field equation \eqref{g eom} for $i = 2$, namely
\begin{multline*}
D_{22} \gamma^{0 \alpha} (g^{-1} \partial_{\alpha} g)^{(2)} + R^{(2)}_g \big( D_{22} (g^{-1} \partial_{\sigma} g)^{(2)} \big)\\
= (g^{-1} X g)^{(2)} - R^{(2)}_g \left( R^{(2)}_g (g^{-1} X g) - (g^{-1} X g)^{(1)} + (g^{-1} X g)^{(3)} + \frac{D_{11} \gamma^{00}}{\sqrt{1+\eta^2}} (\mu^{(1)} - \mu^{(3)}) \right),
\end{multline*}
and compare this to the difference between the field equations in \eqref{g eom} for $i = 1$ and $i = 3$, which can be written as
\begin{multline*}
D_{11} \gamma^{0 \alpha} (g^{-1} \partial_{\alpha} g)^{(1)} - D_{33} \gamma^{0 \alpha} (g^{-1} \partial_{\alpha} g)^{(3)} - D_{22} (g^{-1} \partial_{\sigma} g)^{(2)} + E_g \big( D_{22} (g^{-1} \partial_{\sigma} g)^{(2)} \big)\\
= R^{(2)}_g (g^{-1} X g) - E_g \left( R^{(2)}_g (g^{-1} X g) - (g^{-1} X g)^{(1)} + (g^{-1} X g)^{(3)} + \frac{D_{11} \gamma^{00}}{\sqrt{1+\eta^2}} (\mu^{(1)} - \mu^{(3)}) \right),
\end{multline*}
where we have defined $E_g = 1 + R^{(1)}_g - R^{(3)}_g$. In view of the similarity of some of the terms in the above two equations, it is natural to introduce the following operator
\begin{equation} \label{Q def}
Q = R^{(2)}_g \circ E_g^{-1}.
\end{equation}
Using this we obtain the following equation for the grade 2 part of the dynamics of $g$,
\begin{subequations} \label{IgXg system}
\begin{align}
D_{22} \gamma^{0 \alpha} (g^{-1} \partial_{\alpha} g)^{(2)} &+ Q \big( - D_{11} \gamma^{0 \alpha} (g^{-1} \partial_{\alpha} g)^{(1)} + D_{22} (g^{-1} \partial_{\sigma} g)^{(2)} + D_{33} \gamma^{0 \alpha} (g^{-1} \partial_{\alpha} g)^{(3)} \big) \notag\\
&= (g^{-1} X g)^{(2)} - Q \circ R^{(2)}_g (g^{-1} X g).
\end{align}
Notice that the Lagrange multipliers $\mu^{(1)}$ and $\mu^{(3)}$
are no longer present in this equation. We have made use of
the grade 1 and grade 3 parts of the field equations to eliminate them.
 To solve for $g^{-1} X g$ we will combine this equation with the two fermionic constraints \eqref{C1 C3} which can respectively be rewritten using $\C^{(0)} = - (g^{-1} X g)^{(0)} \simeq 0$ as
\begin{align}
-D_{11} (g^{-1} \partial_{\sigma} g)^{(1)} &\simeq (g^{-1} X g)^{(1)} + R_g^{(1)} \big( (g^{-1} X g)^{(1)} + (g^{-1} X g)^{(2)} + (g^{-1} X g)^{(3)} \big),\\
D_{33} (g^{-1} \partial_{\sigma} g)^{(3)} &\simeq (g^{-1} X g)^{(3)} - R_g^{(3)} \big( (g^{-1} X g)^{(1)} + (g^{-1} X g)^{(2)} + (g^{-1} X g)^{(3)} \big).
\end{align}
\end{subequations}

Now we claim that the system of equations \eqref{IgXg system} can be written in the following matrix form
\begin{equation} \label{IgXg123 bis}
D \left( \!\! \begin{array}{c} - (g^{-1} \partial_{\sigma} g)^{(1)} \\ \gamma^{0 \alpha} (g^{-1} \partial_{\alpha} g)^{(2)} \\ (g^{-1} \partial_{\sigma} g)^{(3)} \end{array} \!\! \right) = K \left( \!\! \begin{array}{c} (g^{-1} X g)^{(1)} \\ (g^{-1} X g)^{(2)} \\ (g^{-1} X g)^{(3)} \end{array} \!\! \right) - \tilde{K} D \left( \!\! \begin{array}{c} - \gamma^{0 \alpha} (g^{-1} \partial_{\alpha} g)^{(1)} \\ (g^{-1} \partial_{\sigma} g)^{(2)} \\ \gamma^{0 \alpha} (g^{-1} \partial_{\alpha} g)^{(3)} \end{array} \!\! \right).
\end{equation}
where the various matrices are defined as
\beqz
K = \left( \!\! \begin{array}{ccc}
1 + R^{(1)}_g & R^{(1)}_g & R^{(1)}_g\\
- Q \circ R^{(2)}_g & 1 - Q \circ R^{(2)}_g & - Q \circ R^{(2)}_g\\
- R^{(3)}_g & - R^{(3)}_g & 1 - R^{(3)}_g
\end{array} \!\! \right),
\qquad
\tilde{K} = \left( \!\! \begin{array}{ccc}
0 & 0 & 0\\
Q & Q & Q\\
0 & 0 & 0
\end{array} \!\! \right).
\eeqz
To solve the system \eqref{IgXg123 bis} for $g^{-1} X g$ we should invert the matrix $K$. Although this matrix has non-commuting entries, it has the (right) Manin matrix property: entries of the same row commute $[K_{ij}, K_{ik}] = 0$ and cross-commutators are equal $[K_{ij}, K_{kl}] = [K_{il}, K_{kj}]$ for all $i, j, k, l$ (in other words the transpose $K^{\top}$ is a usual (left) Manin matrix). In particular, its row ordered determinant is given simply by
\beqz
\text{rdet} \, K = E_g - Q \circ R^{(2)}_g = (E_g - R^{(2)}_g) E^{-1}_g (E_g + R^{(2)}_g).
\eeqz
Now provided this operator is invertible, we can construct
the inverse of $K$. Supposing that $E_g \pm R^{(2)}_g$ are invertible,
it is straightforward to show that
the inverse matrix $K^{-1}$ exists and is given explicitly by
$K^{-1} = \ha (K_+^{-1} + K_-^{-1})$ where the matrices
$K_{\pm}^{-1}$ read
\begin{equation*}
K_{\pm}^{-1} = {\bf 1} - \left( \!\! \begin{array}{ccc}
R^{(1)}_g & R^{(1)}_g & R^{(1)}_g\\
\mp R^{(2)}_g & \mp R^{(2)}_g & \mp R^{(2)}_g\\
- R^{(3)}_g & - R^{(3)}_g & - R^{(3)}_g
\end{array} \!\! \right) \frac{1}{E_g \mp R^{(2)}_g}.
\end{equation*}
Note that multiplication by the inverse of $E_g \mp R^{(2)}_g$ is on the right.
Moreover, one can also show that $K^{-1} \tilde{K} = \ha (K_+^{-1} - K_-^{-1})$.
We can thus invert the above system \eqref{IgXg123 bis} and write
\begin{equation*}
\left( \!\! \begin{array}{c} (g^{-1} X g)^{(1)} \\ (g^{-1} X g)^{(2)} \\ (g^{-1} X g)^{(3)} \end{array} \!\! \right) = \ha (K_+^{-1} + K_-^{-1}) D \left( \!\! \begin{array}{c} - (g^{-1} \partial_{\sigma} g)^{(1)} \\ \gamma^{0 \alpha} (g^{-1} \partial_{\alpha} g)^{(2)} \\ (g^{-1} \partial_{\sigma} g)^{(3)} \end{array} \!\! \right) + \ha (K_+^{-1} - K_-^{-1}) D \left( \!\! \begin{array}{c} - \gamma^{0 \alpha} (g^{-1} \partial_{\alpha} g)^{(1)} \\ (g^{-1} \partial_{\sigma} g)^{(2)} \\ \gamma^{0 \alpha} (g^{-1} \partial_{\alpha} g)^{(3)} \end{array} \!\! \right).
\end{equation*}
Alternatively, introducing the usual combinations $P^{\alpha \beta}_{\pm} = \ha \big(\gamma^{\alpha \beta} \pm \epsilon^{\alpha \beta} \big)$ we can also write this as
\begin{equation} \label{IgXg123}
\left( \!\! \begin{array}{c} (g^{-1} X g)^{(1)} \\ (g^{-1} X g)^{(2)} \\ (g^{-1} X g)^{(3)} \end{array} \!\! \right) = K_+^{-1} D \left( \!\! \begin{array}{c} - P^{0 \alpha}_+ (g^{-1} \partial_{\alpha} g)^{(1)} \\ P^{0 \alpha}_+ (g^{-1} \partial_{\alpha} g)^{(2)} \\ P^{0 \alpha}_+ (g^{-1} \partial_{\alpha} g)^{(3)} \end{array} \!\! \right) + K_-^{-1} D \left( \!\! \begin{array}{c} P^{0 \alpha}_- (g^{-1} \partial_{\alpha} g)^{(1)} \\ P^{0 \alpha}_- (g^{-1} \partial_{\alpha} g)^{(2)} \\ - P^{0 \alpha}_- (g^{-1} \partial_{\alpha} g)^{(3)} \end{array} \!\! \right).
\end{equation}

Finally, by adding the three components of the vector equation
\eqref{IgXg123} and using the constraint $\C^{(0)} = - (g^{-1} X g)^{(0)} \simeq
0$ we obtain the desired equation  
\begin{equation} \label{IgXg final}
g^{-1} X g \simeq \frac{(1 + \eta^2)^2}{2 (1 - \eta^2)} \left( P^{0 \alpha}_+ \frac{1}{1 + \eta \, \tilde{d} \circ R_g} \tilde{d} \, (g^{-1} \partial_{\alpha} g)  + P^{0 \alpha}_- \frac{1}{1 - \eta \, d \circ R_g} d \, (g^{-1} \partial_{\alpha} g) \right),
\end{equation}
where the following combinations of the projectors onto
$\mathfrak{f}^{(i)}$ have been defined:
\begin{equation*}
d = P_1 + \frac{2}{1 - \eta^2} P_2 - P_3, \qquad
\tilde{d} = - P_1 + \frac{2}{1 - \eta^2} P_2 + P_3.
\end{equation*}

\subsection{Deformed Action}

The last step is to plug the relation \eqref{IgXg final} and the constraints into
the functional
\eqref{eqtlvar}.  This means in particular that we need to
compute (see \eqref{constraints d})
\beq \label{hefinal}
h_\epsilon = \lambda^+ \mathcal{T}_+ + \lambda^- \mathcal{T}_-\simeq  \lambda^+ \str(A_+^{(2)} A_+^{(2)}) +
\lambda^- \str(A_-^{(2)} A_-^{(2)})
\eeq
with $A_\pm^{(2)}$ defined in \eqref{def Apm2}.
 Recall that  the fields
$\lambda^\pm$ are related to the worldsheet metric
by equation \eqref{deflambdapm}. At this point, it is useful to introduce
\beq \label{def of J and Jt}
\JJ_\alpha = \frac{1}{1 - \eta R_g \circ \dd} (g^{-1} \partial_\alpha g), \qquad
\Jt_\alpha = \frac{1}{1 + \eta R_g \circ \dt}(g^{-1} \partial_\alpha g).
\eeq
 With such definitions, the equation \eqref{IgXg final} for $g^{-1} X g$
may  be rewritten as
\beq \label{IgXg2}
g^{-1} X g = \frac{(1+\eta^2)^2}{2(1-\eta^2)}
 (P_+^{0\alpha} \dt \Jt_\alpha + P_-^{0\alpha} \dd \JJ_\alpha ).
\eeq
The Lagrangian expressions for $A_\pm^{(2)}$ are then computed from
\eqref{APi to gX} and \eqref{IgXg2}.
The result of this computation is:
\beq
A_+^{(2)}  = - \frac{1+\eta^2}{1-\eta^2} P_-^{0\alpha} \JJ_\alpha^{(2)},
\qquad A_-^{(2)}  = - \frac{1+\eta^2}{1-\eta^2} P_+^{0\alpha} \Jt_\alpha^{(2)}.
\label{aug08d}
\eeq
To bring the action into its final form, we will make use of the following identities,
\begin{align*}
\str \bigl( g^{-1} \partial_\alpha g \dd \JJ_\beta \bigr) &=
\str \bigl( g^{-1} \partial_\beta g \dt \Jt_\alpha \bigr),\\
\str\left(g^{-1} \partial_{(\alpha}  g \dd \JJ_{\beta)} \right) &= \str(\JJ_{(\alpha} \dd \JJ_{\beta)})
= \frac{2}{1-\eta^2} \str(\JJ^{(2)}_\alpha  \JJ^{(2)}_\beta)
=  \frac{2}{1-\eta^2} \str(\Jt^{(2)}_\alpha  \Jt^{(2)}_\beta).
\end{align*}
  They can be proved by using  the antisymmetry of $R$ and the property
$\str(M \dd N)= \str( (\dt M) N)$. We then find on one hand,
\begin{subequations}
 \begin{align}
  {h}_\epsilon  &=  \left(\frac{1+\eta^2}{1-\eta^2}\right)^2\left[
 \lambda^+  P_-^{0\alpha}  P_-^{0\beta}
+ \lambda^-    P_+^{0\alpha}   P_+^{0\beta}\right]
\str (  \JJ_\alpha^{(2)}  \JJ_\beta^{(2)}),\\
&=-\ha \left(\frac{1+\eta^2}{1-\eta^2}\right)^2 \left[
\gamma^{00} \str(\JJ_0^{(2)} \JJ_0^{(2)}) - \gamma^{11} \str(\JJ_1^{(2)} \JJ_1^{(2)})
\right].
 \end{align}
\end{subequations}
On the other hand, we get
\begin{subequations}
 \begin{align}
 \str\left(g^{-1} \partial_\tau g g^{-1}X g\right) &=
\frac{(1+\eta^2)^2}{2(1-\eta^2)}   \str\left(
P_-^{\alpha 0} g^{-1}\partial_\alpha g \dd \JJ_0
+ P_-^{0\alpha} g^{-1}\partial_0 g \dd \JJ_\alpha \right),\\
&= \frac{(1+\eta^2)^2}{2(1-\eta^2)}  P_-^{\alpha\beta}
  \str\left(  g^{-1} \partial_\alpha g  \dd \JJ_\beta \right) - {h}_\epsilon.
 \end{align}
\end{subequations}
As a consequence, the deformed action stemming
from \eqref{eqtlvar} is
\beq \label{PRL Action}
S_\epsilon[g]=-\frac{(1+\eta^2)^2}{2(1-\eta^2)} \int d\sigma d\tau \bigl(  P_-^{\alpha\beta}
  \str\left(  g^{-1} \partial_\alpha g  \dd \JJ_\beta \right) \bigr).
\eeq
This action is the starting point of the analysis carried  in
\cite{Delduc:2013qra}.

 \subsection{Comments on invertibility of $1- \eta R_g \circ d$}
 \label{text inver}

An interesting feature of this computation is that to perform the inverse
Legendre transform and
therefore define the theory at the lagrangian level, we have to take the
inverse of the operators
$1-\eta R_g \circ d$ and $1+\eta R_g \circ \tilde{d}$. This is necessary in order to
invert
the relation between $g^{-1} \partial_\tau g$ and $g^{-1} X g$, as can be seen
from equation \eqref{IgXg final}. And these operators appear in the deformed
action \eqref{PRL Action} via the definitions \eqref{def of J and Jt} of $\JJ_\alpha$ and $\Jt_\alpha$.
It is therefore important to study the invertibility of these operators. This is
discussed in appendix \ref{sec_invR}. Let us briefly summarize the situation here.
The invertibility depends on the choice made for $R$ and must be studied case by case.
It is known \cite{Delduc:2013fga} that the invertibility holds in the compact bosonic sector
regardless of the choice made for $R$. It is also known from the results of
 \cite{Arutyunov:2013ega} that for the choice made in \cite{Delduc:2013qra}, the
operator $1-\eta R_g \circ d$ is not invertible everywhere on the bosonic non-compact
sector. As a consequence, the deformed metric associated with the
action \eqref{PRL Action} exhibits a singularity \cite{Arutyunov:2013ega},
whose meaning is not yet clear 
 (see also \cite{Kameyama:2014bua}}  for a related discussion).  
In appendix \ref{sec_invR}, we study different
choices of $R$ in the bosonic non-compact and fermionic sectors.

\subsection{$\kappa$-symmetry}

At the hamiltonian level, the $\kappa$-symmetry transformations are
generated by
\beqz
g^{-1} \delta g = \left\{ \int_{-\infty}^{\infty} d\sigma
\str(\psi^{(1)} {\mathcal{K}}^{(3)} + \psi^{(3)}
{\mathcal{K}}^{(1)}), g \right\}_\epsilon.
\eeqz
The first class constraints ${\mathcal{K}}^{(1)}$ and ${\mathcal{K}}^{(3)}$
are given by \eqref{Kappa13} while
$\psi^{(1)}$ and $\psi^{(3)}$ are the parameters of this transformation.
This variation can be easily computed by using the expressions \eqref{C1 C3}
of $\C^{(1)}$ and $\C^{(3)}$  and the Poisson brackets \eqref{PB gX}.
One finds  
\beqz
g^{-1} \delta g \simeq - \frac{2 i }{\sqrt{1+\eta^2}}
\left( (1+\eta R_g)[A_-^{(2)}, \psi^{(3)}]_+
+ (1-\eta R_g)[A_+^{(2)}, \psi^{(1)}]_+\right).
\eeqz
The corresponding transformation at the lagrangian level is then
obtained by substituting the lagrangian expressions \eqref{aug08d}
of $A_\pm^{(2)}$
into this result. This leads to  
\beqz
g^{-1} \delta g \simeq \frac{2 i \sqrt{1+\eta^2}}{1-\eta^2} \left(
(1+\eta R_g)[P_+^{0\alpha}
\Jt_\alpha^{(2)} , \psi^{(3)}]_+
+ (1-\eta R_g)[P_-^{0\alpha} \JJ_\alpha^{(2)}, \psi^{(1)}]_+\right).
\eeqz
Note that the variation $g^{-1} \delta g$  does not lie purely in
the odd part of $\mathfrak{psu}(2,2|4)$
contrary to what happens in the underformed case.

\section{Conclusion}

In this article we gave a direct derivation of the integrable $q$-deformation of the $AdS_5 \times S^5$ superstring action. Its tree-level light-cone $S$-matrix in the bosonic sector was determined in \cite{Arutyunov:2013ega} and shown to match the large string tension limit of the $S$-matrix with $q$-deformed centrally-extended $[\mathfrak{psu}(2|2)]^2$ symmetry \cite{Beisert:2008tw,Beisert:2010kk,Hoare:2011wr} for $q$ real.
The relation found in \cite{Arutyunov:2013ega} between the real parameters
$q$ and $\eta$ is exactly as in equation \eqref{q def}. Furthermore, we have
shown that the deformed theory, before gauge-fixing, admits a symmetry
which is the classical analog of the quantum group $U_q(\mathfrak{psu}(2,2|4))$.
These results leave no doubt that the conclusion of \cite{Arutyunov:2013ega}
should extend to the full light-cone $S$-matrix. An interesting duality was also found
in \cite{Arutynov:2014ota, Arutyunov:2014cra}, relating the deformed superstring for
two values of the deformation parameter $\eta$ through mirror duality. It would be
interesting to understand if such a duality admits a classical interpretation using the
hamiltonian framework.

\medskip

An interesting limit of the deformed theory is its ``maximally deformed'' limit
given by $\eta \to 1$, or equivalently $\varkappa \to \infty$ where
$\varkappa = 2 \eta / (1 - \eta^2)$. We have identified this limit at
  the hamiltonian level in subsection \ref{ds5h5}. It is in agrement
with the conjecture we made in \cite{Delduc:2013qra}. Note, however, that
this limit  cannot    be taken straightforwardly at the lagrangian level.
This can already be understood from the relations \eqref{APi to gX}
between $(A,\Pi)$ and $(g,X)$.
Nevertheless, there has been significant progress in understanding
the nature of the geometry corresponding to the bosonic part of the
deformed action   in this limit.
In particular, it was studied in \cite{Hoare:2014pna} using the
parameterization of \cite{Arutyunov:2013ega}, where it was found that the
deformed metric in the limit
$\varkappa \to \infty$ only corresponds to $dS_5 \times H^5$ after applying
some $T$-duality transformations. Furthermore, the same conclusion
was reached more recently in  \cite{Arutyunov:2014cra} but with a different
combination  of $T$-dualities.
 What distinguishes the results of \cite{Hoare:2014pna} and
\cite{Arutyunov:2014cra} is the way the
various fields are scaled in the limit $\varkappa \to \infty$.

We believe that the situation may be clarified from
the hamiltonian perspective. Indeed, it is expected from the hamiltonian
analysis carried out in subsection \ref{ds5h5} that in the limit $\varkappa \to
  \infty$ one
should introduce another field $\hat{g}$ taking values in $PSU^\ast(4|4)$.
This limit and the precise relation
between the fields $g$ and $\hat{g}$ within the present formalism deserve
further study.
It is worth also noting that for some deformed symmetric space $\sigma$-models,
 the maximally deformed limit of
the geometry is relatively simple. This is the case of the $SU(2)/U(1)$ example
considered in \cite{Delduc:2013fga}. But the inspection of other low
dimensional cases suggests that for spheres
$S^{2n}$ and anti-de Sitter spaces $AdS_{2n}$ of even dimension, the
curvature of the maximally deformed background is constant and negative,
respectively positive, without the need of performing any $T$-duality.

 \medskip

One of our original motivations for deforming the $AdS_5 \times S^5$ superstring came from the desire to understand the classical theory which may underly the $q$-deformed $S$-matrix of \cite{Hoare:2011fj, Hoare:2011wr, Hoare:2011nd, Hoare:2012fc, Hoare:2013ysa}. In fact, the linear combination of compatible Poisson brackets used in constructing the deformed theory is very reminiscent of the interpolating nature of this $S$-matrix \cite{Hoare:2011wr}, between the $S$-matrix of the $AdS_5 \times S^5$ superstring in light-cone gauge and the $S$-matrix of the Pohlmeyer reduced theory \cite{Grigoriev:2007bu, Mikhailov:2007xr}. However, the deformation parameter $q$ entering this $S$-matrix is taken to be a root of unity.
Hence a natural question concerns the possibility of constructing a deformation for which $q$ and $\eta$ are complex.
Constructing such a deformation in the present framework would require using a split solution of mCYBE on $\mathfrak{psu}(2,2|4)$.
However, skew-symmetric split solutions of the modified classical Yang-Baxter equation are known to exist mostly for split real forms.
Let us also note that the real form $U_q(\mathfrak{psu}(2,2|4))$ requires $q$ to be real. This problem deserves further investigation.

\medskip

Despite these issues regarding the reality conditions on $q$ and the connection
with Pohlmeyer reduced theory, the authors of
\cite{Hoare:2014pna} considered the bosonic light-cone theory associated
with the deformed $AdS_5 \times S^5$ geometry defined by \eqref{PRL Action}, for the standard choice of $R$, in the limit where $\eta = i$, or equivalently $\varkappa = i$.
More precisely, the part of the full geometry relevant for this computation contains the deformed metric and the $B$-field. When taking $\varkappa=i$, the deformed metric
remains real but the $B$-field becomes imaginary. Interestingly, it was found
in \cite{Hoare:2014pna} that, when discarding
the imaginary $B$-field, the expansion up to quartic
order in certain fields of the bosonic light-cone action
associated with the deformed metric
agrees with that of the Pohlmeyer reduction of the
$AdS_5 \times S^5$ superstring \cite{Hoare:2009fs}.
Furthermore, it was shown that when truncating the deformation
to $AdS_3 \times S^3$, this agreement even holds \cite{Hoare:2009fs}
to all orders for the bosonic fields but also for the quadratic
fermionic terms. Note that there is in this case no need to
discard the imaginary $B$-field as it vanishes for the deformed
$AdS_3 \times S^3$ geometry. It would be interesting
to understand this within the present hamiltonian formalism.
Some progress in this direction was
made in \cite{Hollowood:2014fha} at the level of the
generalised sine-Gordon theories.

\medskip

The deformed action \eqref{PRL Action} is a generalisation of the Yang-Baxter $\sigma$-model action \cite{Klimcik:2008eq}. In particular, it is also characterised by a non-split solution of the modified classical Yang-Baxter equation. It is possible to extend this action to the case where the $R$-matrix involved is a solution of the classical Yang-Baxter equation (CYBE). Such an action was studied in \cite{Kawaguchi:2014qwa,
Kawaguchi:2014fca, Matsumoto:2014nra, Matsumoto:2014gwa,Crichigno:2014ipa}. It was shown
in particular that the $\gamma$-deformation
\cite{Lunin:2005jy,Frolov:2005ty,Frolov:2005dj} falls within this class of deformations. It would be interesting to derive such deformed actions from first principle in the spirit of the present article.

\medskip

All these remarks lead in fact to the same and therefore important
question of understanding how the choice of $R$-matrix used in the construction
affects the deformation. Indeed, the deformed action depends on a non-split
solution
of the modified classical Yang-Baxter equation, which in turn can be
associated with any choice of system of positive roots of $\mathfrak{psu}(2,2|4)$.
Let us emphasise that the deformed geometry may typically depend on this choice.
An important and related question is whether the resulting deformed geometry defines
a background of Type IIB supergravity. This remains an open question.

\paragraph{Acknowledgements.} We would like to thank A. Anabalon, S. Frolov, B. Hoare,
A. Tseytlin and S. van Tongeren for useful discussions. This work is partially supported by the
program PICS 6412 DIGEST of CNRS.

\appendix

\section{The real Lie algebras $\f$ and $\p\f$}
\label{app: psu}

\subsection{The Lie superalgebras $\mathfrak{gl}(4|4)$, $\mathfrak{sl}(4|4)$ and $\mathfrak{psl}(4|4)$}

Let $E_{ab}$ be the standard basis of generators for the Lie superalgebra $\mathfrak{gl}(4|4)$ with defining $\mathbb{Z}_2$-graded commutation relations
\begin{equation} \label{gl44 def}
[E_{ab}, E_{cd}] = \delta_{bc} E_{ad} - (-1)^{|E_{ab}| |E_{cd}|} \delta_{ad} E_{cb}.
\end{equation}
The parity of $E_{ab}$ is defined as $|E_{ab}| = |a| + |b| \in \mathbb{Z}_2$ where $|a| = 0$ if $a \leq 4$ and $|a| = 1$ if $a \geq 5$. Let $\overline{\mathfrak{H}}$ be the span of the generators $E_{aa}$ for $1 \leq a \leq 8$. We denote by $\mathfrak{gl}(4|4)^{[k]}$, $k = 0, 1$ the subspaces of $\mathfrak{gl}(4|4)$ spanned by all $E_{ab}$ with $|E_{ab}| = k$.

The subalgebra $\mathfrak{sl}(4|4)$ is spanned by all generators $E_{ab}$ with $a \neq b$ together with the following combinations of the Cartan generators
\begin{gather}
H_1 =
E_{11} - E_{22}, \qquad
H_2 =
E_{22} - E_{33}, \qquad
H_3 =
E_{33} - E_{44}, \qquad
H_4 =
E_{44} + E_{55}, \notag\\
H_5 =
E_{55} - E_{66}, \qquad
H_6 =
E_{66} - E_{77}, \qquad
H_7 =
E_{77} - E_{88}. \label{Cartan gen}
\end{gather}
Introduce the corresponding subspaces $\mathfrak{sl}(4|4)^{[k]} = \mathfrak{sl}(4|4) \cap \mathfrak{gl}(4|4)^{[k]}$ for $k = 0, 1$.
The generator $I = \sum_{a=1}^8 E_{aa} \in \mathfrak{sl}(4|4)$ is central in $\mathfrak{sl}(4|4)$ and the quotient by the ideal spanned by $I$ defines the Lie superalgebra $\mathfrak{psl}(4|4)$.

In the fundamental representation of $\mathfrak{gl}(4|4)$, $E_{ab}$ is represented by the $8 \times 8$ matrix $e_{ab}$ whose only non-zero entry is a $1$ in the $a^{\rm th}$ row and $b^{\rm th}$ column. We equip $\mathfrak{gl}(4|4)$ with a non-degenerate bilinear graded-symmetric invariant form $(\cdot, \cdot) : \mathfrak{gl}(4|4) \times \mathfrak{gl}(4|4) \to \mathbb{C}$ defined by taking the supertrace of the product in the fundamental representation. It is given in the basis $E_{ab}$ by
\begin{equation} \label{ip Eab}
(E_{ab}, E_{cd}) = \str(e_{ab} e_{cd}) = \delta_{bc} \delta_{ad} (-1)^{|a|}.
\end{equation}

\paragraph{$\mathbb{Z}_4$-automorphism.} Recall that $\mathfrak{gl}(4|4)$ is equipped with an automorphism $\Omega : \mathfrak{gl}(4|4) \to \mathfrak{gl}(4|4)$ of order 4. Letting $t$ be the permutation $(12)(34)(56)(78)$, it can be defined on the generators $E_{ab}$ as
\begin{equation} \label{Omega def}
\Omega(E_{ab}) = (-1)^{a + b + 1 + |a| (1 - |b|)} E_{t(b) t(a)}.
\end{equation}
Let $\mathfrak{gl}(4|4)^{(j)}$, $0 \leq j \leq 3$ denote the eigenspace of $\Omega$ with eigenvalue $i^j$, so that for $X^{(j)} \in \mathfrak{gl}(4|4)^{(j)}$,
\begin{equation} \label{propOmega}
\Omega(X^{(j)}) = i^j X^{(j)}.
\end{equation}
Noting $\Omega^2(E_{ab}) = (-1)^{|E_{ab}|} E_{ab}$ it follows that $\mathfrak{gl}(4|4)^{(0)}$, $\mathfrak{gl}(4|4)^{(2)}$ are both subspaces of $\mathfrak{gl}(4|4)^{[0]}$ and $\mathfrak{gl}(4|4)^{(1)}$, $\mathfrak{gl}(4|4)^{(3)}$ are subspaces of $\mathfrak{gl}(4|4)^{[1]}$. The automorphism $\Omega$ preserves the subalgebra $\mathfrak{sl}(4|4)$ and since $\Omega(I) = - I$ it also induces an automorphism on $\mathfrak{psl}(4|4)$.

\paragraph{Root system.} With respect to the Cartan subalgebra $\overline{\mathfrak{H}}$, the root space of $\mathfrak{gl}(4|4)$ (and $\mathfrak{sl}(4|4)$) is given by $\Phi = \{ \epsilon_a - \epsilon_b \,|\, 1 \leq a \neq b \leq 8 \}$ where $\epsilon_a = (-1)^{|a|} (E_{aa} , \cdot)$.
The root $\epsilon_a - \epsilon_b$ is called even if $|a| = |b|$ and odd if $|a| \neq |b|$.
A positive system of roots in $\Phi$ is uniquely specified by a permutation $(a_1, \ldots, a_8)$ of $(1, \ldots, 8)$ and is given as $\Phi^+ = \{ \epsilon_{a_{\mu}} - \epsilon_{a_{\nu}} \,|\, 1 \leq \mu < \nu \leq 8 \}$.
The corresponding set of simple roots then reads $\Delta = \{ \alpha_{\mu} \,|\, 1 \leq {\mu} \leq 7 \}$ where we have defined $\alpha_{\mu} = \epsilon_{a_{\mu}} - \epsilon_{a_{\mu+1}}$.

Given any root $\alpha \in \Phi$ we denote the corresponding root vector as $E^{\alpha}$, which has the property that
\begin{equation} \label{H com E}
[H, E^{\alpha}] = \alpha(H) E^{\alpha}
\end{equation}
for any Cartan generator $H$. In particular, $E^{\epsilon_a - \epsilon_b}$ for $a \neq b$ is proportional to $E_{ab}$. In order to fix the normalisation we will use equation \eqref{ip Eab}. Specifically, given any positive root $\alpha = \epsilon_a - \epsilon_b \in \Phi^+$ we define
\begin{equation*}
E^{\alpha} = E^{\epsilon_a - \epsilon_b} = E_{ab}, \qquad
E^{- \alpha} = E^{\epsilon_b - \epsilon_a} = (-1)^{|a|} E_{ba}.
\end{equation*}
It then follows that for any $\alpha \in \Phi^+$, equation \eqref{ip Eab} takes the form
\begin{equation} \label{ip Eab 2}
\big( E^{\alpha}, E^{\beta} \big) = \delta_{\alpha, - \beta}.
\end{equation}
For each positive root $\alpha \in \Phi^+$ we then define the Cartan element
\begin{equation*}
H^{\alpha} = [E^{\alpha}, E^{-\alpha}].
\end{equation*}
Explicitly, for a positive root of the form $\alpha = \epsilon_a - \epsilon_b \in \Phi^+$ we have $H^{\epsilon_a - \epsilon_b} = (-1)^{|a|} E_{aa} - (-1)^{|b|} E_{bb}$, with the property that $(H^{\alpha}, H) = \alpha(H)$ for any Cartan element $H$.
A useful basis of the Cartan subalgebra $\mathfrak{H}$ of $\mathfrak{sl}(4|4)$ is given by the generators $H^{\mu} = H^{\alpha_{\mu}}$ for each simple root $\alpha_{\mu} \in \Delta$, $\mu = 1, \ldots, 7$. We also define the symmetric bilinear pairing on roots as $(\alpha, \beta) = \alpha(H^{\beta}) = (H^{\alpha}, H^{\beta})$ for any $\alpha, \beta \in \Phi$.

\paragraph{Cartan matrix.} The symmetrised Cartan matrix $(B_{\mu \nu})_{\mu, \nu = 1}^7$ is defined as $B_{\mu \nu} = \alpha_{\mu}(H^{\nu})$.
It is singular since $\alpha_{\mu}(I) = 0$ for $1 \leq \mu \leq 7$ using the fact that $I = \sum_{a=1}^8 E_{aa} \in \mathfrak{H}$ is central in $\mathfrak{sl}(4|4)$. Writing the latter as $I = \sum_{\nu = 1}^7 x_{\nu} H^{\nu}$ for some $x_{\nu} \in \mathbb{Z}$ we have
\begin{equation} \label{B singular}
\sum_{\nu = 1}^7 B_{\mu \nu} x_{\nu} = 0.
\end{equation}

In order to deal with the Cartan matrix being singular we enlarge the Cartan subalgebra $\mathfrak{H}$ of $\mathfrak{sl}(4|4)$ to $\overline{\mathfrak{H}}$ by adding the extra generator $H^8 = \ha \sum_{i=1}^8 (-1)^{|a|} E_{aa}$, which amounts to working instead with $\mathfrak{gl}(4|4)$. The symmetrised Cartan matrix may now be extended using the commutation relations \eqref{H com E} for $H^8$ to obtain the extended symmetrised Cartan matrix $\big( \overline{B}_{ab} \big)_{a, b = 1}^8$. Specifically, we have
\begin{equation*}
[H^a, E^{\pm \alpha_{\mu}}] = \pm \overline{B}_{\mu a} E^{\pm \alpha_{\mu}}, \qquad
\overline{B}_{\mu a} = \alpha_{\mu}(H^a)
\end{equation*}
for $1 \leq \mu \leq 7$ and $1 \leq a \leq 8$. The remaining components $\overline{B}_{8 \nu}$ with $1 \leq \nu \leq 7$ are then defined by symmetry and we set $\overline{B}_{88} = 0$. Explicitly, we have
\begin{equation} \label{barB def}
\big( \overline{B}_{ab} \big)_{a, b = 1}^8 = \left(\!\!
\begin{array}{cc}
B_{\mu \nu} & \alpha_{\mu}(H^8)\\
\alpha_{\nu}(H^8) & 0
\end{array}
\!\!\right).
\end{equation}
Since the generators $H^a$, $a = 1, \ldots, 8$ form a basis of $\overline{\mathfrak{H}}$ it is clear that this matrix is non-degenerate. Its inverse can also be written explicitly as
\begin{equation} \label{barB inv}
\big( \overline{B}^{-1}_{ab} \big)_{a, b = 1}^8 = \left(\!\!
\begin{array}{cc}
Y_{\mu \nu} & \omega^{-1} x_{\mu}\\
\omega^{-1} x_{\nu} & 0
\end{array}
\!\!\right)
\end{equation}
where $\omega = \sum_{\mu = 1}^7 x_{\mu} \alpha_{\mu}(H^8)$ and the matrix $(Y_{\mu \nu})_{\mu, \nu = 1}^7$ satisfies the following relation
\begin{equation} \label{BX prod}
\sum_{\rho = 1}^7 B_{\mu \rho} Y_{\rho \nu} + \omega^{-1} x_{\nu} \alpha_{\mu}(H^8) = \delta_{\mu \nu}.
\end{equation}
Of course, the matrix $(Y_{\mu \nu})_{\mu, \nu = 1}^7$ is also singular since we have $\sum_{\nu = 1}^7 Y_{\mu \nu} \alpha_{\nu}(H^8) = 0$.

\paragraph{Tensor Casimir.}
The tensor Casimir $C^{\mathfrak{gl}}_{\1\2}$ of $\mathfrak{gl}(4|4)$, with the property that $(C^{\mathfrak{gl}}_{\1\2}, X_{\2})_{\2} = X_{\1}$ for any $X \in \mathfrak{gl}(4|4)$, reads
\begin{equation*}
C^{\mathfrak{gl}}_{\1\2} = \sum_{a, b = 1}^8 (-1)^{|b|} E_{ab} \otimes E_{ba}.
\end{equation*}
It will be convenient for us to also rewrite this Casimir in terms of Cartan-Weyl generators, which can be done as follows. The generators $E_{ab}$ with $a \neq b$ already correspond to root generators since $E_{ab} = E^{\epsilon_a - \epsilon_b}$. As for the Cartan part of $C^{\mathfrak{gl}}_{\1\2}$, it can also be re-expressed in terms of the basis $H^a$, $a = 1, \dots, 8$ of $\overline{\mathfrak{H}}$ by using the extended symmetrised Cartan matrix, namely
\begin{equation} \label{gl Casimir}
C^{\mathfrak{gl}}_{\1\2} = \sum_{a, b = 1}^8 \overline{B}_{ab}^{-1} H^a \otimes H^b + \sum_{\alpha > 0} \big( (-1)^{|E^{\alpha}|} E^{\alpha} \otimes E^{-\alpha} + E^{-\alpha} \otimes E^{\alpha} \big).
\end{equation}
In fact, by using the explicit form \eqref{barB inv} for the inverse of the extended symmetrised Cartan matrix, we may rewrite \eqref{gl Casimir} more explicitly as $C^{\mathfrak{gl}}_{\1\2} = C_{\1\2} + \kappa^{-1} \big( I \otimes H^8 + H^8 \otimes I \big)$
where
\begin{equation} \label{psl Casimir}
C_{\1\2} = \sum_{\mu, \nu = 1}^7 Y_{\mu \nu} H^{\mu} \otimes H^{\nu} + \sum_{\alpha > 0} \big( (-1)^{|E^{\alpha}|} E^{\alpha} \otimes E^{-\alpha} + E^{-\alpha} \otimes E^{\alpha} \big).
\end{equation}
This is the Casimir for $\mathfrak{psl}(4|4)$.

\paragraph{Examples.} The standard positive system corresponds to the permutation $(1, 2, 3, 4, 5, 6, 7, 8)$. In this case, the root vectors associated with positive roots are just the generators $E_{ab}$ with $a < b$ and the Cartan generators $H^{\mu}$ are identified with the generators $H_{\mu}$ defined in \eqref{Cartan gen}. The corresponding Dynkin diagram and extended symmetrised Cartan matrix are
\begin{equation*}
\begin{tikzpicture}[baseline =-5,scale=.6]
\draw[thick] (1,0) -- (7,0);
\foreach \x in {1,2,3,4,5,6,7}
\filldraw[fill=white, thick] (\x,0) circle (2mm);
\draw[thick] (4,0)++(-.15,-.15) -- ++(.3,.3);\draw[thick] (4,0)++(.15,-.15) -- ++(-.3,.3);
\end{tikzpicture}
\qquad\qquad
\big( \overline{B}_{ab} \big)_{a, b = 1}^8 = \left(
\begin{array}{cccccccc}
2 & -1 &  &  &  &  &  & \\
-1 & 2 & -1 &  &  &  &  & \\
 & -1 & 2 & -1 &  &  &  & \\
 &  & -1 & 0 & 1 &  &  & 1\\
 &  &  & 1 & -2 & 1 &  & \\
 &  &  &  & 1 & -2 & 1 & \\
 &  &  &  &  & 1 & -2 & \\
 &  &  & 1 &  &  &  & 0
\end{array}
\right),
\end{equation*}
where a node
\begin{tikzpicture}[baseline =-3,scale=.6] \filldraw[fill=white, thick] (0,0) circle (2mm);\end{tikzpicture}
(resp.
\begin{tikzpicture}[baseline =-3,scale=.6] \filldraw[fill=white, thick] (0,0) circle (2mm);\draw[thick] (0,0)++(-.15,-.15) -- ++(.3,.3);\draw[thick] (0,0)++(.15,-.15) -- ++(-.3,.3);\end{tikzpicture})
represents an even (resp. odd) simple root.

By contrast, the positive system defined by the permutation
$(5, 6, 1, 2, 3, 4, 7, 8)$ corresponds to the ``Beauty''
Dynkin diagram \cite{Beisert:2003yb} and has the following
extended symmetrised Cartan matrix
\begin{equation*}
\begin{tikzpicture}[baseline =-5,scale=.6]
\draw[thick] (1,0) -- (7,0);
\foreach \x in {1,2,3,4,5,6,7}
\filldraw[fill=white, thick] (\x,0) circle (2mm);
\draw[thick] (2,0)++(-.15,-.15) -- ++(.3,.3);\draw[thick] (2,0)++(.15,-.15) -- ++(-.3,.3);
\draw[thick] (6,0)++(-.15,-.15) -- ++(.3,.3);\draw[thick] (6,0)++(.15,-.15) -- ++(-.3,.3);
\end{tikzpicture}
\qquad\qquad
\big( \overline{B}_{ab} \big)_{a, b = 1}^8 = \left(
\begin{array}{cccccccc}
-2 & 1 &  &  &  &  &  & \\
1 & 0 & -1 &  &  &  &  & -1\\
 & -1 & 2 & -1 &  &  &  & \\
 &  & -1 & 2 & -1 &  &  & \\
 &  &  & -1 & 2 & -1 &  & \\
 &  &  &  & -1 & 0 & 1 & 1\\
 &  &  &  &  & 1 & -2 & \\
 & -1 &  &  &  & 1 &  & 0
\end{array}
\right).
\end{equation*}

\subsection{The real forms $\mathfrak{su}(2,2|4)$ and $\mathfrak{psu}(2,2|4)$}

The real form $\mathfrak{su}(2,2|4)$ of $\mathfrak{sl}(4|4)$ is defined as follows. Let $s$ be the function on $\{1, \ldots, 8 \}$ such that $s(a) = 1$ if $a = 3, 4$ and $s(a) = 0$ otherwise. We introduce an anti-linear involutive automorphism $\tau$ of $\mathfrak{sl}(4|4)$ by defining it on generators as
\begin{equation} \label{psi reality}
\tau(H^{\mu}) = - H^{\mu}, \qquad
\tau(E_{ab}) = - (-1)^{s(a) + s(b)} i^{- |E_{ab}|} E_{ba},
\end{equation}
where $1 \leq \mu \leq 7$, $1 \leq a \neq b \leq 8$ and then extending it to all of $\mathfrak{sl}(4|4)$ by anti-linearity. It has the properties
\begin{equation*}
\tau(\lambda X + \mu Y) = \overline{\lambda} \, \tau(X) + \overline{\mu} \, \tau(Y), \qquad
\tau^2 = 1, \qquad
\tau\big( [X, Y] \big) = \big[ \tau(X), \tau(Y) \big],
\end{equation*}
for any $\lambda, \mu \in \mathbb{C}$ and $X, Y \in \mathfrak{sl}(4|4)$.
The real Lie superalgebra $\mathfrak{su}(2,2|4)$ can now be defined as the subalgebra of $\mathfrak{sl}(4|4)$ consisting of $\tau$-invariant elements. A basis of $\mathfrak{su}(2,2|4)$ is given by
\begin{equation} \label{real basis}
T^{\mu} = i H^{\mu}, \qquad
B^{\alpha} = i \big( E^{\alpha} - \tau(E^{\alpha}) \big), \qquad
C^{\alpha} = E^{\alpha} + \tau(E^{\alpha}),
\end{equation}
where $1 \leq \mu \leq 7$ and $\alpha \in \Phi^+$. We define $\mathfrak{su}(2,2|4)^{[k]} = \mathfrak{su}(2,2|4) \cap \mathfrak{sl}(4|4)^{[k]}$, $k = 0, 1$.
A basis for $\mathfrak{su}(2,2|4)^{[0]}$ is then given by $T^{\mu}$ and $B^{\alpha}$, $C^{\alpha}$ for any even positive root $\alpha \in \Phi^+$, while a basis for $\mathfrak{su}(2,2|4)^{[1]}$ consists of all remaining generators $B^{\alpha}$ and $C^{\alpha}$ with odd positive root $\alpha \in \Phi^+$.
Finally, the real Lie superalgebra $\mathfrak{psu}(2,2|4)$ is obtained as a quotient of $\mathfrak{su}(2,2|4)$ by the ideal spanned by $I$. It also admits a $\mathbb{Z}_2$-grading $\mathfrak{psu}(2,2|4)^{[k]}$ with $k = 0, 1$ induced from that of $\mathfrak{su}(2,2|4)$.

\paragraph{$\mathbb{Z}_4$-automorphism.} It follows from the definitions \eqref{Omega def} of the $\mathbb{Z}_4$-automorphism $\Omega$ and \eqref{psi reality} of the anti-linear automorphism $\tau$ that $\Omega \circ \tau(E_{ab}) = (-1)^{|E_{ab}|} \tau \circ \Omega(E_{ab})$. Combining this with the definition \eqref{propOmega} of the eigenspaces of $\Omega$, it follows that $\tau$ preserves each of the graded components $\mathfrak{sl}(4|4)^{(j)}$, $0 \leq j \leq 3$. We may therefore define the real subspaces $\mathfrak{su}(2,2|4)^{(j)} = \mathfrak{su}(2,2|4) \cap \mathfrak{sl}(4|4)^{(j)}$ so that for any $X^{(j)} \in \mathfrak{su}(2,2|4)^{(j)}$ we have $\tau(X^{(j)}) = X^{(j)}$. Note from the property \eqref{propOmega}, however, that $\Omega$ does not preserve the odd subspaces $\mathfrak{su}(2,2|4)^{(1)}$ and $\mathfrak{su}(2,2|4)^{(3)}$.

\subsection{The Grassmann envelopes $\f$ and $\p\f$}

Let $\Gr^{\mathbb{C}}$ be a Grassmann algebra, namely an algebra over $\mathbb{C}$ generated by anti-commuting variables $\xi_a$, $a = 1, \ldots, N$. A general $\xi \in \Gr^{\mathbb{C}}$ is a finite linear combination of products of the $\xi_a$. Denote by $(\Gr^{\mathbb{C}})^{[k]}$, $k = 0, 1$ the subspaces of sums containing only products of an even (respectively odd) number of generators $\xi_a$.

We equip $\Gr^{\mathbb{C}}$ with an anti-linear involution $\xi \mapsto \xi^{\ast}$ for any $\xi \in \Gr^{\mathbb{C}}$
satisfying
\begin{equation*}
(c \, \xi)^{\ast} = \overline{c} \, \xi^{\ast}, \qquad
(\xi^{\ast})^{\ast} = \xi, \qquad
(\xi \zeta)^{\ast} = \xi^{\ast} \zeta^{\ast},
\end{equation*}
for $\xi, \zeta \in \Gr^{\mathbb{C}}$ and $c \in \mathbb{C}$. Define the real Grassmann algebra $\Gr$ as the subalgebra of elements $\xi \in \Gr^{\mathbb{C}}$ such that $\xi^{\ast} = \xi$. Correspondingly, the real Grassmann envelope of $\mathfrak{su}(2,2|4)$ is defined as
\begin{equation*}
\f = \big( \Gr \otimes \mathfrak{su}(2,2|4) \big)^{[0]} = \Gr^{[0]} \otimes \mathfrak{su}(2,2|4)^{[0]} \oplus \Gr^{[1]} \otimes \mathfrak{su}(2,2|4)^{[1]}.
\end{equation*}
This is an ordinary Lie algebra with $\mathbb{Z}_2$-grading $\f^{[k]} = \Gr^{[k]} \otimes \mathfrak{su}(2,2|4)^{[k]}$ for $k = 0, 1$. We denote its complexification by $\f^{\mathbb{C}} = \big( \Gr^{\mathbb{C}} \otimes \mathfrak{sl}(4|4) \big)^{[0]}$. If we extend $\tau$ to an anti-linear homomorphism of $\f^{\mathbb{C}}$ by setting $\tau(\xi \otimes X) = \xi^{\ast} \otimes \tau(X)$ for $\xi \in \Gr^{\mathbb{C}}$ and $X \in \mathfrak{sl}(4|4)$, then $\f$ becomes the fixed point subalgebra of $\f^{\mathbb{C}}$, namely $\f = \{ x \in \f^{\mathbb{C}} \, | \, \tau(x) = x \}$.
We introduce also the Lie algebra $\p\f = \big( \Gr \otimes \mathfrak{psu}(2,2|4) \big)^{[0]}$ with $\mathbb{Z}_2$-graded subspaces $\p\f^{[k]} = \Gr^{[k]} \otimes \mathfrak{psu}(2,2|4)^{[k]}$ for $k = 0, 1$.

The graded-symmetric bilinear form \eqref{ip Eab} on $\mathfrak{sl}(4|4)$ extends to a symmetric bilinear form $(\cdot, \cdot) : \f^{\mathbb{C}} \times \f^{\mathbb{C}} \to \Gr^{\mathbb{C}}$ on the Grassmann envelope $\f^{\mathbb{C}}$ by letting $(\xi \otimes X, \zeta \otimes Y) = (-1)^{|\zeta| |X|} \xi \zeta (X, Y)$ for any $\xi, \zeta \in \Gr^{\mathbb{C}}$ and $X, Y \in \mathfrak{sl}(4|4)$. Restricting to the real Grassmann envelope $\f$ we obtain a symmetric bilinear form $(\cdot, \cdot) : \f \times \f \to \Gr$.

In the fundamental representation of $\mathfrak{su}(2,2|4)$, the Lie algebra $\f$ consists of block diagonal even supermatrices
\begin{equation*}
M = \left( \!\! \begin{array}{cc} a & \psi \\ \chi & b \end{array} \!\! \right)
\end{equation*}
where $a, b$ are $4 \times 4$ matrices with entries in $\Gr^{[0]}$ and $\psi, \chi$ are $4 \times 4$ matrices with entries in $\Gr^{[1]}$, and satisfying the relation
\begin{equation*}
(M^{\ast})^{st} S + S M = 0, \qquad
M^{st} = \left( \!\! \begin{array}{cc} a^{\sf T} & - \chi^{\sf T} \\ \psi^{\sf T} & b^{\sf T} \end{array} \!\! \right), \qquad
S = \text{diag}({\bf 1}_2, -{\bf 1}_2, i {\bf 1}_4).
\end{equation*}
Here $a^{\sf T}$ denotes the transpose of a $4 \times 4$ matrix with entries in $\Gr$. We therefore have
\beq \label{taufondamental}
\tau(M) = - S^{-1} (M^{\ast})^{st} S
\eeq
in the fundamental representation of $\mathfrak{su}(2,2|4)$.

Likewise, the $\mathbb{Z}_4$-automorphism of $\mathfrak{sl}(4|4)$, defined on generators in \eqref{Omega def}, can be expressed in the fundamental representation as
\begin{equation} \label{auto Omega}
\Omega(M)=-{\mathbf K}^{-1} M^{st} {\mathbf K},\qquad
{\mathbf K}= \text{diag}(\mathsf{k}, \mathsf{k}, \mathsf{k}, \mathsf{k}),\qquad
\mathsf{k}=
\left(\begin{array}{cc}0&-1\cr 1&0\end{array}\right).
\end{equation}

\section{Non-split $R$-matrix}
\label{app: nsrmatrix}
Following the conventions laid out in appendix \ref{app: psu}, we fix a positive system $\Phi^+$ of roots in $\Phi$. Let $\mathfrak{B}$ denote the corresponding Borel subalgebra of $\mathfrak{sl}(4|4)$ which by definition is spanned by the Cartan generators $H^{\mu} \in \mathfrak{H}$ along with the positive root vectors $E^{\alpha}$, $\alpha \in \Phi^+$. It is clear from \eqref{psi reality} that $\tau$ sends $\mathfrak{B}$ into its opposite Borel subalgebra $\tau(\mathfrak{B})$, spanned by $H^{\mu}$ and $E^{-\alpha}$ for $\alpha \in \Phi^+$.

\paragraph{Conjugate Borel subalgebras.} We define a subalgebra $\b$ of $\f^{\mathbb{C}}$ by letting
\begin{equation} \label{Borel sub}
\b = \big( \Gr^{\mathbb{C}} \otimes \mathfrak{B} \big)^{[0]}.
\end{equation}
Explicitly, $\b$ is spanned by elements of the form $\xi \otimes H^{\mu}$ with $|\xi| = 0$ and $\xi \otimes E^{\alpha}$, $\alpha \in \Phi^+$ with $|\xi| = |E^{\alpha}|$.
Since $\mathfrak{B}$ and $\tau(\mathfrak{B})$ are opposite Borel subalgebras of $\mathfrak{sl}(4|4)$ it follows that
\begin{equation} \label{tau b}
\b + \tau(\b) = \f^{\mathbb{C}}.
\end{equation}
Let $\h = \b \cap \tau(\b)$ which   is spanned by elements
$\xi \otimes H^{\mu}$ with $|\xi| = 0$ and   define
 the nilpotent subalgebra $\n = [\b, \b]$. Then $\b = \h \oplus \n$ and
we have the vector space decomposition $\f^{\mathbb{C}} = \n \oplus \tau(\b)$.

\paragraph{Decomposition of $\f^{\mathbb{C}}$ relative to $\f$.}
Let $\h_0 = \{ h \in \h \,|\, \tau(h) = - h \}$. Using the first relation in
\eqref{psi reality} it follows that $\h_0$ is the linear span over
the real Grassmann envelope of the Cartan generators in \eqref{Cartan gen}.
That is, $\h_0$ consists of elements of the form
$\xi_\mu \otimes H^{\mu}$ where $\xi_\mu \in \Gr^{[0]}$
and $1 \leq \mu \leq 7$. Now we claim that as vector spaces,
\begin{equation} \label{decomposition}
\f^{\CC} = \f \oplus \h_0 \oplus \n.
\end{equation}
Indeed, using the decomposition $\f^{\CC} = \n \oplus \tau(\b)$ we may write any $x \in \f^{\CC}$ as $x = n + h + X$ where $n \in \n$, $X \in \tau(\n)$ and $h \in \tau(\h)$. On the other hand we have
\begin{equation*}
X + h = \big( (X + \ha h) + \tau(X + \ha h) \big) + \ha \big(h - \tau(h)\big) - \tau(X) \in \f \oplus \h_0 \oplus \n,
\end{equation*}
so that $x \in \f^{\CC}$ can be written as a sum in $\f \oplus \h_0 \oplus \n$. Such a decomposition is clearly unique since the three subalgebras $\f$, $\h_0$ and $\n$ have pairwise trivial intersection.

\paragraph{Non-split $R$-matrix.} We introduce a $\Gr$-linear operator $R : \f \to \f$ defined relative to a choice of subalgebra $\b$ in \eqref{Borel sub} as follows. First note that any $x \in \f$ can be written uniquely in the form $x = \frac{i}{2} (b - \tau(b))$ for some $b \in \h_0 \oplus \n$. Indeed, such an expression can be obtained by decomposing $- i x \in \f^{\CC}$ relative to \eqref{decomposition} as $- i x = y + b$ with $y \in \f$ and $b \in \h_0 \oplus \n$. Moreover, it is unique since if $x = \frac{i}{2} (c - \tau(c))$ for some $c \in \h_0 \oplus \n$ then it follows that $b - c \in \f$ and therefore $b = c$. We now define $R$ as
\begin{equation} \label{R def}
R\big( i (b - \tau(b)) \big) = b + \tau(b)
\end{equation}
for all $b \in \h_0 \oplus \n$. It is straightforward to check that this is a skew-symmetric `non-split' solution of the modified classical Yang-Baxter equation, that is to say it satisfies $(R x, y) = - (x, R y)$ and
\begin{equation} \label{mCYBE}
[R x, R y] - R \big( [R x, y] + [x, R y] \big) = [x, y],
\end{equation}
for any $x, y \in \f$. Indeed, writing $x = i (b - \tau(b))$ and $y = i (c - \tau(c))$ for $b, c \in \h_0 \oplus \n$ we have
\begin{equation*}
\big( R \big(i (b - \tau(b)) \big), i (c - \tau(c)) \big) + \big( i (b - \tau(b)), R \big(i (c - \tau(c)) \big) \big) = 2 i (b, c) - 2 i (\tau(b), \tau(c)),
\end{equation*}
which vanishes since $b, c \in \h_0 \oplus \n$. Moreover, for each term in \eqref{mCYBE} we find
\begin{align*}
\big[ R x, R y \big] &= [b, c] + \tau\big( [b, c] \big) + [b, \tau(c) ] + \tau \big( [b, \tau(c)] \big),\\
R \big( [R x, y] + [x, R y] \big) &= 2 [b, c] + 2 \tau\big( [b, c] \big),\\
[x, y] &= - [b, c] - \tau\big( [b, c] \big) + [b, \tau(c)] + \tau\big( [b, \tau(c)] \big).
\end{align*}
The $R$-matrix also has the property that $(R \mp i) : \f \to \f^{\mathbb{C}}$ project onto the positive and negative Borel subalgebras $\b$ and $\tau(\b)$ of $\f^{\mathbb{C}}$, respectively. More specifically, for any $x \in \f$ we have
\begin{equation} \label{R pm i}
(R - i) x \in \h_0 \oplus \n, \qquad
(R + i) x \in \h_0 \oplus \tau(\n).
\end{equation}

Given a particular choice of Borel subalgebra $\mathfrak{B}$, the $R$-matrix \eqref{R def} may be written explicitly as follows. We first introduce an $\mathbb{R}$-linear operator $R : \mathfrak{su}(2,2|4) \to \mathfrak{su}(2,2|4)$ by defining it on the basis generators \eqref{real basis} of $\mathfrak{su}(2,2|4)$. For the Cartan generators we set $R(T^{\mu}) = 0$. Next, for every positive root $\alpha \in \Phi^+$ we define
\begin{equation} \label{R def st}
R(B^{\alpha}) = C^{\alpha}, \qquad R(C^{\alpha}) = - B^{\alpha}.
\end{equation}
These expressions can be obtained from an analogous formula to \eqref{R def} but for $b \in \mathfrak{B}$.
In particular, this is a non-split solution of the super mCYBE, namely
\begin{equation} \label{smCYBE}
[RX, RY] - R\big( [RX, Y] + [X, RY] \big) = [X, Y].
\end{equation}
Extending $R$ to the real Grassmann envelope by letting $R(\xi \otimes X) = \xi \otimes R(X)$ for any $\xi \in \Gr$ and $X \in \mathfrak{su}(2,2|4)$, we obtain a skew-symmetric operator $R : \f \to \f$ satisfying the usual mCYBE \eqref{mCYBE}.

\section{$q$-Poisson-Serre relations} \label{app: qSerre}

In this appendix, we prove the standard $q$-Poisson-Serre relations
  \eqref{all stand qPS} and the non-standard one \eqref{all non stand qPS}.

\subsection{First set of standard $q$-Poisson-Serre relations}

We start by proving that
\beq
\{Q^E_{\alpha_\nu} , Q^E_{\alpha_\mu} \}_\epsilon=0  \qquad \mbox{when} \qquad
(\alpha_\nu , \alpha_\mu)=0.  \label{ma sqps 1}
\eeq
The charge $Q^E_{\alpha_\nu}$ defined by \eqref{QHE def}, with $\mathfrak{J}^E_{\alpha_\nu }(\sigma)$ given by \eqref{JH JE}, is merely the integral of the density $e_{\alpha_\nu}(\sigma) e^{-\gamma \chi_{\alpha_\nu}(\sigma)} e^{\gamma \chi_{\alpha_\nu}(-\infty)}$. Thus, when computing the Poisson bracket of $Q^E_{\alpha_\nu}$ and $Q^E_{\alpha_\mu}$, we have three different kinds of terms. It is however clear that they all vanish.  Indeed, the first kind of terms comes from   Poisson brackets of $\chi_{\alpha_\nu}$ with $\chi_{\alpha_\mu}$. They vanish by
 using the definition \eqref{chi def}  of $\chi _{\alpha_\nu}$ and the
Poisson bracket \eqref{jh jh pb}. The second kind of terms comes from  Poisson brackets of
 $e_{\alpha_\nu}(\sigma)$ with $e^{-\gamma \chi_{\alpha_\mu}(\sigma')}$ and
those with $(\nu, \sigma)$ and $(\mu,\sigma')$ flipped. However, the
result \eqref{chi e pb} indicates that these Poisson brackets are both
proportional to the element $B_{\nu\mu}$ of the symmetrized Cartan matrix,
and therefore vanish in the case at hand. Finally, the last kind of term originates
from the Poisson bracket of  $e_{\alpha_\nu}(\sigma)$ with $e_{\alpha_\mu}(\sigma')$.
This Poisson bracket has to be extracted from the Poisson bracket of $X(\sigma)$
with $X(\sigma')$ and may be read off from
\eqref{PB eX} by using \eqref{another def X}. But more generally, the
Poisson bracket \eqref{PB gX c} of $X$ with itself  is
just a Kirillov-Kostant Poisson bracket associated with $\mathfrak{psu}(2,2|4)$. It is therefore clear that  the Poisson bracket of $e_{\alpha_\nu}(\sigma)$ with $e_{\alpha_\mu}(\sigma')$ vanishes in the present case. This ends the proof of \eqref{ma sqps 1}.

\subsection{Second set of standard $q$-Poisson-Serre relations}

Next, we prove that
\beq \label{realqpsm21a}
\{  \{ Q^E_{\alpha_{\mu}} , Q^E_{\alpha_{\mu+1}} \}_{q \, \epsilon} ,
Q^E_{\alpha_{\mu+1}} \}_{q \, \epsilon} = 0
\eeq
with $\alpha_\mu < \alpha_\mu + \alpha_{\mu +1} < \alpha_{\mu +1}$.
The relation
\begin{equation} \label{qPSrel2}
\{  Q^E_{\alpha_{\mu}} , \{   Q^E_{\alpha_{\mu}}   , Q^E_{\alpha_{\mu+1}}
 \}_{q \, \epsilon}  \}_{q \, \epsilon} = 0
\end{equation}
is proved in a similar way.

\paragraph{Intermediate results.} We begin by listing some of the properties that will be used
in proving \eqref{realqpsm21a}. These properties all hold when $\alpha_\mu + \alpha_\nu$ is a root.

We define $N_{\alpha_\mu, \alpha_{\nu}}$ by
\beq \label{def of N}
[E^{\alpha_\mu}, E^{\alpha_{\nu}}] = N_{\alpha_\mu, \alpha_{\nu}} E^{\alpha_\mu +\alpha_{\nu} }.
\eeq
Starting from the relation \eqref{PB eX} one can then show that
\beq  \label{may27y}
N_{\alpha_\mu, \alpha_\nu} \{ e_{\alpha_\mu}(\sigma) , e_{\alpha_\nu}(\sigma') \}_\epsilon
= 2 i  (-1)^{|E^{\alpha_{\mu} }| |E^{\alpha_\nu}|} (\alpha_\mu,\alpha_\nu) e_{\alpha_\mu + \alpha_\nu}(\sigma) \delta_{\sigma \sigma'}.
\eeq

Let $\alpha$ and $\beta$ be two positive roots. It immediately follows from
the generalisation of the Poisson bracket \eqref{chi e pb} to arbitrary positive roots $\alpha$ and $\beta$ that
\beq \label{may27z}
 \{ e^{-\gamma ( \chi_{\alpha}(\sigma) -\chi_\alpha(-\infty) ) } , e_{\beta}(\sigma') \}_\epsilon
= i   \gamma (\alpha ,\beta)   e^{-\gamma ( \chi_{\alpha}(\sigma) -\chi_\alpha(-\infty) )} e_{\beta}(\sigma')  \theta_{\sigma \sigma'},
\eeq
where $\theta_{\sigma \sigma'} = \ha (\epsilon_{\sigma \sigma'} +1)$ is the Heaviside step function.

The results \eqref{may27y} and \eqref{may27z} may be combined to prove that
\begin{subequations}
\beq \label{J J alone}
\{ \mathfrak{J}^E_{\alpha_{\nu}}(\sigma), \mathfrak{J}^E_{\alpha_{\mu}}(\sigma')\}_\epsilon
=  - 2 i   N_{\alpha_\mu, \alpha_\nu} (\alpha_\nu,\alpha_\mu) \mathfrak{J}^E_{\alpha_\nu +\alpha_\mu}(\sigma) \delta_{\sigma \sigma'} +   i \gamma (\alpha_\nu,\alpha_\mu) \mathfrak{J}^E_{\alpha_\nu}(\sigma)  \mathfrak{J}^E_{\alpha_\mu}(\sigma') \epsilon_{\sigma \sigma'},
\eeq
when $ N_{\alpha_\mu, \alpha_\nu} \neq 0$.  We have made use of the definition \eqref{JH JE}  of $\mathfrak{J}^E_\alpha(\sigma)$, the (anti)-symmetry property $(-1)^{|E^{\alpha_{\nu } }| |E^{\alpha_\mu}|} N_{\alpha_{\nu },\alpha_\mu}  = - N_{\alpha_\mu, \alpha_{\nu}}$ and  the relation  $N^2_{\alpha_\mu, \alpha_{\nu}} = 1$.
In particular, it follows that
\begin{multline} \label{eq101}
\{ \mathfrak{J}^E_{\alpha_{\nu}}(\sigma), \mathfrak{J}^E_{\alpha_{\mu}}(\sigma')\}_\epsilon
+ i \gamma (\alpha_{\nu},\alpha_\mu) \mathfrak{J}^E_{\alpha_{\nu }} (\sigma) \mathfrak{J}^E_{\alpha_{\mu}}(\sigma') \\
=- 2 i   N_{\alpha_\mu, \alpha_\nu} (\alpha_\nu,\alpha_\mu) \mathfrak{J}^E_{\alpha_\nu +\alpha_\mu}(\sigma) \delta_{\sigma \sigma'}
+ 2 i \gamma (\alpha_\nu,\alpha_\mu) \mathfrak{J}^E_{\alpha_\nu}(\sigma)  \mathfrak{J}^E_{\alpha_\mu}(\sigma') \theta_{\sigma \sigma'}.
\end{multline}
\end{subequations}

We will also make use of the following results:
\begin{subequations} \label{KKOrPS}
\begin{align}
\{e_{\alpha_{\mu}}(\sigma), e_{ \alpha_\mu }(\sigma')\}_\epsilon &=0, \label{may27zb} \\
\{e_{\alpha_{\mu+1}}(\sigma), e_{\alpha_{\mu} + \alpha_{\mu+1} }(\sigma')\}_\epsilon &=0. \label{eip1eiip1pb}
\end{align}
\end{subequations}
The first relation comes from the fact that $2 \alpha_{\mu}$ is not a root. The second
relation is a consequence of \eqref{PB eX} and the ordinary Serre relation $[E^{\alpha_{\mu+1}}, [E^{\alpha_{\mu+1}}, E^{\alpha_\mu}]] = 0$.

A consequence of \eqref{may27zb} and \eqref{may27z} is that we have
\beq \label{pb Jmup Jmup}
 \{ \mathfrak{J}^E_{\alpha_{\mu}} (\sigma), \mathfrak{J}^E_{\alpha_{\mu}}(\sigma') \}_\epsilon=  i \gamma (\alpha_{\mu} , \alpha_{\mu}) \mathfrak{J}^E_{\alpha_{\mu}} (\sigma) \mathfrak{J}^E_{
\alpha_{\mu} }(\sigma') \epsilon_{\sigma \sigma'}.
 \eeq

\paragraph{$q$-Poisson-Serre relation.}
Following the approach of \cite{Delduc:2013fga}, we first show that
\beq \label{qpsm21a}
\{ Q^E_{\alpha_{\mu} }, Q^E_{\alpha_{\mu+1}} \}_{q \, \epsilon} =
- 2 i N_{ \alpha_{\mu+1},\alpha_\mu} (\alpha_{\mu+1} , \alpha_\mu)
Q^E_{\alpha_{\mu}+\alpha_{\mu+1}}.
\eeq
This is simply done by integrating   \eqref{eq101} in the case $\nu = \mu+1$ and remembering that (see \eqref{res21mas}) the density $\mathfrak{Q}^E_{\alpha_{\mu} + \alpha_{\mu+1}}(\sigma)$ is defined by
\beq \label{remindQamu}
\mathfrak{Q}^E_{\alpha_{\mu} + \alpha_{\mu+1}}(\sigma) = \mathfrak{J}^E_{\alpha_{\mu} + \alpha_{\mu+1}}(\sigma) - \gamma \, N_{\alpha_{\mu}, \alpha_{\mu+1}} \mathfrak{J}^E_{\alpha_{\mu+1}}(\sigma) \int_{-\infty}^{\sigma} d\sigma' \mathfrak{J}^E_{\alpha_{\mu}}(\sigma').
\eeq
With the help of \eqref{qpsm21a}, proving the $q$-Poisson-Serre relation
\eqref{realqpsm21a}  means  showing that
\beq \label{another way of qPS2}
\{ Q^E_{\alpha_{\mu}+\alpha_{\mu+1}} ,
Q^E_{\alpha_{\mu+1}} \}_{q \, \epsilon} = 0.
\eeq
This  is equivalent to proving $\{ Q^E_{\alpha_{\mu+1}},
Q^E_{\alpha_{\mu}+\alpha_{\mu+1}}
 \}_{q^{-1} \, \epsilon} = 0$,
or, in other words,
\beq \label{integrand qPS2bis}
\int_{-\infty}^\infty d\sigma  \int_{-\infty}^\infty d\sigma'
\Bigl( \{ \mathfrak{ J}^E_{\alpha_{\mu+1}}(\sigma), \mathfrak{Q}^E_{\alpha_{\mu} + \alpha_{\mu+1}}(\sigma') \}_\epsilon   + i \gamma (\alpha_{\mu+1},\alpha_\mu+\alpha_{\mu+1}) \mathfrak{ J}^E_{\alpha_{\mu+1}}(\sigma) \mathfrak{Q}^E_{\alpha_\mu  + \alpha_{\mu+1}}(\sigma') \Bigr)=0.
\eeq
Let us first evaluate $\{ \mathfrak{ J}^E_{\alpha_{\mu+1}}(\sigma), \mathfrak{Q}^E_{\alpha_\mu + \alpha_{\mu+1}}(\sigma') \}_\epsilon$.
  Using   the definition \eqref{JH JE}, the property \eqref{may27z} for $\alpha =\alpha_\mu$ and $\beta = \alpha_{\mu+1} + \alpha_\mu$ and the result \eqref{eip1eiip1pb} leads to
\beq \label{may27za}
 \{ \mathfrak{J}^E_{\alpha_{\mu+1}} (\sigma), \mathfrak{J}^E_{
\alpha_\mu + \alpha_{\mu+1}}(\sigma') \}_\epsilon=  i \gamma (\alpha_\mu + \alpha_{\mu+1}, \alpha_{\mu+1}) \mathfrak{J}^E_{\alpha_{\mu+1}} (\sigma) \mathfrak{J}^E_{
\alpha_\mu + \alpha_{\mu+1}}(\sigma') \epsilon_{\sigma \sigma'}.
\eeq
We then obtain
\begin{align*}
\{ \mathfrak{ J}^E_{\alpha_{\mu+1}}(\sigma), \mathfrak{Q}^E_{\alpha_\mu +\alpha_{\mu+1}}(\sigma') \}_\epsilon
&=   i \gamma (\alpha_\mu + \alpha_{\mu+1}, \alpha_{\mu+1}) \mathfrak{J}^E_{\alpha_{\mu+1}} (\sigma) \mathfrak{J}^E_{\alpha_\mu +\alpha_{\mu+1}}(\sigma') \epsilon_{\sigma \sigma'} \nonumber \\
&- \gamma \, N_{\alpha_\mu, \alpha_{\mu+1}}   \{ \mathfrak{ J}^E_{\alpha_{\mu+1}}(\sigma),
 \mathfrak{J}^E_{\alpha_{\mu+1}}(\sigma') \}_\epsilon \int_{-\infty}^{\sigma'} d\sigma'' \mathfrak{J}^E_{\alpha_\mu}(\sigma'') \nonumber \\
 &  - \gamma \, N_{\alpha_\mu, \alpha_{\mu+1}} (-1)^{ |E^{\alpha_{\mu+1}}|}
  \mathfrak{J}^E_{\alpha_{\mu+1}}(\sigma') \int_{-\infty}^{\sigma'} d\sigma''
  \{ \mathfrak{ J}^E_{\alpha_{\mu+1}}(\sigma), \mathfrak{J}^E_{\alpha_\mu}(\sigma'')\}_\epsilon.
\end{align*}
The complete integrand in \eqref{integrand qPS2bis} may then be written as
 \begin{align*}
   &i \gamma (\alpha_\mu + \alpha_{\mu+1}, \alpha_{\mu+1}) \mathfrak{J}^E_{\alpha_{\mu+1}} (\sigma) \mathfrak{J}^E_{\alpha_\mu +\alpha_{\mu+1}}(\sigma') \epsilon_{\sigma \sigma'}\\
&
- i \gamma^2 \, N_{\alpha_\mu, \alpha_{\mu+1}}   (\alpha_{\mu+1},\alpha_{\mu+1}) \mathfrak{J}^E_{\alpha_{\mu+1}}(\sigma)  \mathfrak{J}^E_{\alpha_{\mu+1}}(\sigma') \epsilon_{\sigma \sigma'}
  \int_{-\infty}^{\sigma'} d\sigma'' \mathfrak{J}^E_{\alpha_\mu}(\sigma'')\\
&  + (-1)^{|E^{\alpha_{\mu+1}}| } 2 i \gamma
(\alpha_{\mu+1},\alpha_\mu)  \mathfrak{J}^E_{\alpha_{\mu+1}}(\sigma') \mathfrak{J}^E_{\alpha_\mu +\alpha_{\mu+1}}(\sigma) \theta_{\sigma'\sigma}\\
& - (-1)^{ |E^{\alpha_{\mu+1}}|} i \gamma^2 N_{\alpha_\mu, \alpha_{\mu+1} }  (\alpha_{\mu+1},\alpha_\mu)
\mathfrak{J}^E_{\alpha_{\mu+1}}(\sigma') \mathfrak{J}^E_{\alpha_{\mu+1}}(\sigma) \int_{-\infty}^{\sigma'} d\sigma''
\mathfrak{J}^E_{\alpha_\mu}(\sigma'')\epsilon_{\sigma \sigma''}\\
   &  + i\gamma (\alpha_{\mu+1},\alpha_\mu+\alpha_{\mu+1}) \mathfrak{ J}^E_{\alpha_{\mu+1}}(\sigma) \mathfrak{ J}^E_{\alpha_\mu + \alpha_{\mu+1}}(\sigma')\\
& - i \gamma^2  N_{\alpha_\mu,\alpha_{\mu+1}} (\alpha_{\mu+1}, \alpha_\mu + \alpha_{\mu+1})
\mathfrak{ J}^E_{\alpha_{\mu+1}}(\sigma)   \mathfrak{ J}^E_{\alpha_{\mu+1}}(\sigma') \int_{-\infty}^{\sigma'} d\sigma'' \mathfrak{ J}^E_{\alpha_\mu}(\sigma''),
\end{align*}
where we have successively used \eqref{remindQamu},  \eqref{may27za}, \eqref{J J alone}
 and \eqref{pb Jmup Jmup}. Adding the terms linear in $\gamma$ on the one hand and those in $\gamma^2$ on the other hand, we find that both sums are proportional to
 \beqz
 (\alpha_\mu + \alpha_{\mu +1}, \alpha_{\mu+1}) + (-1)^{|E^{\alpha_{\mu+1} } |}
 (\alpha_{\mu  }, \alpha_{\mu+1}).
 \eeqz
However,  the value  of this coefficient is:
\begin{alignat}{2}
( \alpha_{\mu +1}, \alpha_{\mu+1}) +  2
 (\alpha_{\mu }, \alpha_{\mu+1}) &= 0, \qquad & \text{for} \; &E^{\alpha_{\mu +1} } \; \text{even}. \\
(  \alpha_{\mu +1}, \alpha_{\mu+1}) &= 0, \qquad & \text{for}\; &E^{\alpha_{\mu +1} } \; \text{odd}.
\end{alignat}
This shows that
$\{ Q^E_{\alpha_{\mu}+\alpha_{\mu+1}} ,
Q^E_{\alpha_{\mu+1}} \}_{q \, \epsilon}  = 0$.

 \subsection{Non-standard $q$-Poisson-Serre relation}

In an analogous way, one can also check the
non-standard $q$-Poisson-Serre relation
\eqref{all non stand qPS}, namely
 \beq \label{appendix non stand qPS}
\{ \{ Q^E_{\alpha_{\mu}}, Q^E_{\alpha_{\mu-1}} \}_{q \, \epsilon} ,
\{ Q^E_{\alpha_{\mu}}, Q^E_{\alpha_{\mu+1}} \}_{q \, \epsilon} \}_\epsilon =0.
\eeq
We will not give full details but just sketch the proof. When computing the left hand side of this relation, one typically gets multiple integrals of terms that are linear, quadratic, cubic and quartic in $\mathfrak{ J}^E_{\alpha_\rho}$ and which contain products of Heaviside step functions. It is clear that the linear term vanishes. This is so because $\alpha_{\mu-1} + 2 \alpha_\mu + \alpha_{\mu+1}$ is not a root. One can show that all other multiple integrals vanish. Let us illustrate this on one type of cubic term and on the quartic term.

 The computation leads to a cubic term proportional to
 \beqz
 \int_{-\infty}^\infty d\sigma \int_{-\infty}^\infty d\sigma' \int_{-\infty}^\infty d\sigma''
 \mathfrak{ J}^E_{\alpha_{\mu-1}}(\sigma')   \mathfrak{ J}^E_{\alpha_{\mu} + \alpha_{\mu+1}}(\sigma)  \mathfrak{ J}^E_{\alpha_{\mu}}(\sigma'')  \bigl(   \theta_{\sigma \sigma''} \theta_{\sigma' \sigma''} -   \theta_{\sigma \sigma'} \theta_{\sigma' \sigma''}   - \theta_{\sigma' \sigma}  \theta_{\sigma \sigma''}   \bigr).
 \eeqz
The appearance of the first product of Heaviside functions means that the domain
 of integration corresponds to $\sigma > \sigma''$ and $\sigma'>\sigma''$. The two other products with the minus sign correspond to the domain $\{\sigma > \sigma' > \sigma''\} \cup \{\sigma'>\sigma>\sigma''\}$. Therefore, the two domains coincide and this cubic term vanishes.

The quartic term is proportional to
\beq  \label{maiy 29}
 \int_{-\infty}^\infty d\sigma \int_{-\infty}^\infty d\sigma' \int_{-\infty}^\infty d\sigma''
  \int_{-\infty}^\infty d\sigma'''
 \mathfrak{ J}^E_{\alpha_{\mu+1}}(\sigma)   \mathfrak{ J}^E_{\alpha_{\mu-1}}(\sigma')
  \mathfrak{ J}^E_{\alpha_{\mu}}(\sigma'')    \mathfrak{ J}^E_{\alpha_{\mu}}(\sigma''')
      \theta_{\sigma\sigma''} \theta_{\sigma' \sigma''}   \theta_{\sigma' \sigma'''}
    \theta_{\sigma \sigma'''}  .
 \eeq
  The product of Heaviside functions in \eqref{maiy 29} is symmetric in the exchange of $\sigma''$ and $\sigma'''$ while the product $\mathfrak{ J}^E_{\alpha_{\mu}}(\sigma'')    \mathfrak{ J}^E_{\alpha_{\mu}}(\sigma''')$ is
 antisymmetric since $E^{\alpha_\mu}$ is odd. Therefore, the quartic contribution vanishes as well.

\paragraph{Comment on literature.}  Let us note that the set of
defining relations for (quantum) superalgebras are sometimes
written differently in the literature
(see for instance  \cite{Yamane_1991a,Zhang:2011ym}).
  Therefore, for completeness, we will also prove that
\begin{equation} \label{june16}
\{ Q^E_{\alpha_{\mu}}, \{ Q^E_{\alpha_{\mu-1}}, \{
 Q^E_{\alpha_{\mu}} ,
Q^E_{\alpha_{\mu+1}}   \}_{q \, \epsilon}
 \}_{q \, \epsilon}   \}_\epsilon =0
\end{equation}
by showing that
\beq \label{twoNSqPSequal}
\{Q^E_{\alpha_{\mu}}, \{ Q^E_{\alpha_{\mu-1}}, Q^E_{\alpha_\mu +\alpha_{\mu+1}}
 \}_{q \, \epsilon}
 \}_\epsilon = \{ \{ Q^E_{\alpha_{\mu}} , Q^E_{\alpha_{\mu - 1}} \}_{q \, \epsilon},
Q^E_{\alpha_\mu +\alpha_{\mu+1}}   \}_\epsilon.
\eeq
To do this, let us start with the left hand side of this equality. By using
the definition \eqref{XdefofqPS} of the $q$-bracket and the
Jacobi identity, one gets
\begin{multline} \label{ma27a}
\{Q^E_{\alpha_{\mu}}, \{ Q^E_{\alpha_{\mu-1}}, Q^E_{\alpha_\mu +\alpha_{\mu+1}}
 \}_{q \, \epsilon}
 \}_\epsilon =
\{ Q^E_{\alpha_{\mu -1}}, \{ Q^E_{\alpha_\mu},
Q^E_{\alpha_\mu+\alpha_{\mu+1}} \}_\epsilon \}_\epsilon \\
+ \{ \{ Q^E_{\alpha_\mu} , Q^E_{\alpha_{\mu-1}} \}_\epsilon ,
Q^E_{\alpha_\mu +\alpha_{\mu+1}}  \}_\epsilon
- i \gamma (\alpha_{\mu -1} , \alpha_\mu )
\{ Q^E_{\alpha_\mu} , Q^E_{\alpha_{\mu -1} }
Q^E_{\alpha_\mu + \alpha_{\mu+1}}    \}_\epsilon.
\end{multline}
We then use the $q$-Poisson-Serre relation \eqref{qPSrel2}, which can
be written as
$\{ Q^E_{\alpha_{\mu}}, Q^E_{\alpha_{\mu} + \alpha_{\mu + 1}}  \}_{q \, \epsilon} = 0$
using \eqref{qpsm21a}, to rewrite the first term on the right
hand side of \eqref{ma27a}. This leads to
\begin{multline} \label{ma27b}
\{Q^E_{\alpha_{\mu}}, \{ Q^E_{\alpha_{\mu-1}}, Q^E_{\alpha_\mu+\alpha_{\mu+1}}
 \}_{q \, \epsilon}
 \}_\epsilon =
 \{ \{ Q^E_{\alpha_\mu} , Q^E_{\alpha_{\mu-1}} \}_\epsilon ,
Q^E_{\alpha_\mu +\alpha_{\mu+1}}  \}_\epsilon \\
+ i \gamma (\alpha_\mu, \alpha_{\mu+1} )
\{ Q^E_{\alpha_{\mu -1} }, Q^E_{\alpha_{\mu} }
 Q^E_{\alpha_\mu + \alpha_{\mu+1} }
 \}_\epsilon
- i \gamma (\alpha_{\mu -1} , \alpha_\mu )
\{ Q^E_{\alpha_\mu}, Q^E_{\alpha_{\mu -1} }
Q^E_{\alpha_\mu + \alpha_{\mu+1}}
 \}_\epsilon.
\end{multline}
Finally, since $ (\alpha_\mu, \alpha_{\mu+1} )  = -  (\alpha_\mu, \alpha_{\mu-1} )$,
the last two terms in the right hand side of \eqref{ma27b} combine
together and give
$  - i \gamma (\alpha_{\mu -1} , \alpha_\mu ) \{ Q^E_{\alpha_\mu}
Q^E_{\alpha_{\mu -1} },
 Q^E_{\alpha_\mu  +\alpha_{\mu+1}}
  \}_\epsilon$.  One then recognizes the right hand side of
\eqref{twoNSqPSequal}. Note that we have used many times that
the parity of $E^{\alpha_\mu}$ is odd. Thus equation \eqref{june16} coincides with equation \eqref{appendix non stand qPS}.

\section{On the invertibility of $1- \eta R_g \circ d$} \label{sec_invR}

We are interested in discussing the invertibility of the linear operator ${\cal O}=1- \eta R_g \circ d$
acting on the Lie algebra $\f$ when
$|\eta| <1$. Recalling the $\mathbb{Z}_2$-grading of the Lie algebra $\f$ from appendix \ref{app: psu}, we denote by
 $P_{[0]} =P_0 + P_2$ and $P_{[1]} =P_1 + P_3$ the projectors on each
 graded components $\f^{[0]}$ and $\f^{[1]}$.
The operator ${\cal O}$ is invertible if and only if its two ``diagonal'' blocks
\beq
{\cal O}_0 = P_{[0]} (1- \eta R_g \circ d) P_{[0]} \qquad
{\cal O}_1 = P_{[1]}  (1- \eta R_g \circ  d )P_{[1]}\label{blocs}
\eeq
are invertible on $\mathfrak{f}^{[0]}$ and $\mathfrak{f}^{[1]}$ respectively.
Moreover, the group element $g$ in (\ref{blocs}) can be restricted to the even subgroup $SU(2,2)\times SU(4)$. In this case, $\text{Ad}\, g$ respects the  $\mathbb{Z}_2$-grading. We will only consider $R$-matrices which also respect the $\mathbb{Z}_2$-grading. Therefore, in the cases considered below, the operator $R_g=\text{Ad}\, g^{-1}\circ R\circ \text{Ad}\, g$ respects the $\mathbb{Z}_2$-grading. Because of this, the operators ${\cal O}_0$ and ${\cal O}_1$ may be rewritten as
\beq
{\cal O}_0 = 1- \varkappa R_g \circ P_2 \qquad
{\cal O}_1 =1- \eta R_g \circ  (P_1 -P_3) ,\label{sblocs}
\eeq
considered as linear operators
acting respectively on $\mathfrak{f}^{[0]}$ and $\mathfrak{f}^{[1]}$.
In this appendix we make use of the notation and parametrisation in \cite{Arutyunov:2013ega}. In particular we have introduced $\varkappa = 2\eta/(1-\eta^2)$.

\subsection{Bosonic sector}

If we restrict attention to deformations of the non-linear $\sigma$-model on the bosonic symmetric space $\frac{SU(2,2)\times SU(4)}{SO(1,4)\times SO(5)}\equiv AdS_5\times S^5$ then only the operator ${\cal O}_0$ is present. The latter was computed in \cite{Arutyunov:2013ega} for a standard choice of $R$ and with an element $g$ which parameterises the coset $AdS_5\times S^5$. It is non-invertible for a particular value of a radial parameter of $AdS_5$ called $\rho$. The singularity takes place at $\rho = 1/\varkappa$ and affects only the deformed metric on the non-compact factor $AdS_5$. Indeed, a general proof of the invertibility of  ${\cal O}_0$ in the case of a compact symmetric space is given in \cite{Delduc:2013fga}.

One might hope that modifying the operator $R$ could improve the situation.
Let us discuss this point in the case of the Lie superalgebra $\mathfrak{su}(2,2)$.
Denote by $\hat R$ the standard antisymmetric non-split solution of
mCYBE acting on $\mathfrak{su}(2,2)$. Let us consider a
permutation $\mathbb{P}$ of $4$ objects, and the corresponding $4\times 4$
matrix ${\cal P}_{ij}=\delta_{i \mathbb{P}(j)}$. We may then construct another
solution of mCYBE as
\beqz
\hat{R}^{\mathbb{P}}=\,\mbox{Ad}\,{\cal P}^{-1} \circ \hat{R} \circ \mbox{Ad}\,{\cal P}.
\eeqz
The reality condition satisfied by any element $M \in \mathfrak{su}(2,2)$
reads $M^{\dag} H + H M = 0$, where $H=\mbox{diag}(1,1,-1,-1)$. If the permuted matrix
$H_{\mathbb{P}}={\cal P}^{-1}H{\cal P}$ coincides with $H$, up to an overall sign,
then the matrix $\hat{R}^{\mathbb{P}}$ leads
to the same deformation of $AdS_5$ as $\hat{R}$ does.
This is so because ${\cal P}$ belongs to $SU(2,2)$, after a possible rescaling by a phase,
and the deformed actions associated with $\hat{R}$ and
$\hat{R}^{\mathbb{P}}$ are related by
\beqz
S_{\hat{R}^{\mathbb{P}}}[g] = S_{\hat{R}}[{{\cal P}} g].
\eeqz
There are therefore only
two permutations which lead to operators $\hat{R}^{\mathbb{P}}$ that are inequivalent to $\hat R$. They are
\beqz
{\mathbb{P}}_1=\left(\begin{array}{cccc}1&2&3&4\cr 1&3&2&4\end{array}\right)
\quad\mbox{and}\quad {\mathbb{P}}_2=\left(\begin{array}{cccc}1&2&3&4\cr 1&3&4&2
\end{array}\right).
\eeqz
By contrast, permutations do not make any difference in the case of the deformation of $S^5$.

Below we give the metric and the $B$-field
associated with the choices $\hat R$, $\hat{R}^{{\mathbb{P}}_1}$
and $\hat{R}^{{\mathbb{P}}_2}$. To fix notations, the restriction to the bosonic non-compact sector of the deformed Lagrangian corresponding to the action \eqref{PRL Action}, is written as
\beqz
  - \ha  (1+\varkappa^2)
\gamma^{\alpha\beta} \partial_\alpha X^M \partial_\beta X^N G_{MN}
+\ha  (1+\varkappa^2)      \,  \epsilon^{\alpha\beta} \partial_\alpha
X^M \partial_\beta X^N B_{MN}.
\eeqz
For convenience, we start by recalling the results of \cite{Arutyunov:2013ega}.
The coordinates $X^M$ used to describe $AdS_5$ are $(t,\rho,\zeta,\psi_1,\psi_2)$.
The metric and the $B$-field associated with the standard choice
$\hat{R}$ take the form
\begin{gather*}
G^{\hat{R}}_{tt} = -\frac{1+\rho^2}{1-\varkappa ^2 \rho ^2 },\qquad
G^{\hat{R}}_{\rho\rho} =  \frac{1}{\left( 1 + \rho ^2 \right) \left(1-\varkappa ^2
   \rho ^2 \right)},
   \qquad
    G^{\hat{R}}_{\zeta \zeta}  =   \frac{\rho^2}{ 1+ \varkappa ^2 \rho ^4 \sin ^2{\zeta} },\\
    G^{\hat{R}}_{\psi_1\psi_1} =  \frac{\rho ^2 \cos ^2{\zeta }}{1+\varkappa ^2 \rho ^4
   \sin ^2{\zeta }},\qquad
 G^{\hat{R}}_{\psi_2\psi_2} = \rho ^2 \sin ^2{\zeta },\\
 B^{\hat{R}}_{\rho t}= \frac{1}{\varkappa} \partial_\rho \log(1- \varkappa ^2 \rho ^2 ),
\qquad
 B^{\hat{R}}_{\psi_1\zeta}=    \varkappa \frac{\rho ^4 \sin (2 \zeta )  }{ 1+
   \varkappa ^2 \rho ^4 \sin ^2{\zeta}  }.
 \end{gather*}
To write down the metrics and $B$-fields associated with the choices of
$R$-matrices $\hat{R}^{\mathbb{P}_1}$ and $\hat{R}^{\mathbb{P}_2}$
more succinctly, we introduce the following functions
\begin{gather*}
f(\rho,\zeta) =1+\varkappa ^2+\varkappa ^2 \rho ^2 \cos ^2{\zeta }, \quad
s(\rho,\zeta)=1-\varkappa^2\rho^2(1+\rho^2\cos^2\zeta)\sin^2\zeta
,\\
h(\rho,\zeta) =1+\varkappa ^2 \left(1+\rho ^2\right)+\varkappa ^2 \rho
   ^2 \left(1+\rho ^2\right) \cos ^2{\zeta }.
\end{gather*}

For the choice of $R$-matrix $\hat{R}^{{\mathbb{P}}_1}$, the non-zero
components of the metric read
\begin{gather*}
G^{{\mathbb{P}}_1}_{tt} = -\frac{ \left(1+\rho ^2\right)}{s(\rho ,\zeta )} ,\qquad
G^{\mathbb{P}_1}_{\rho\rho}=\frac{1+\varkappa^2\sin^2\zeta-\varkappa^2\rho^2(1+\rho^2)\cos^2\zeta\sin^2\zeta}{(1+\rho^2)f(\rho,\zeta)s(\rho,\zeta)}
, \\
G^{\mathbb{P}_1}_{\zeta\zeta}=\frac{\rho^2(1+\varkappa^2(1+\rho^2)\cos^2\zeta-\varkappa^2\rho^2\sin^4\zeta)}{f(\rho,\zeta)s(\rho,\zeta)}
,\\
   G^{{\mathbb{P}}_1}_{\psi_1\psi_1} = \frac{\rho ^2 \cos ^2{\zeta }}{f(\rho ,\zeta )} ,\qquad
 G^{{\mathbb{P}}_1}_{\psi_2\psi_2} = \rho ^2 \sin ^2{\zeta } , \qquad
G^{\mathbb{P}_1}_{\rho\zeta}=\frac{\varkappa^2\rho(1+\rho^2\sin^2\zeta)\cos\zeta\sin\zeta}{f(\rho,\zeta)s(\rho,\zeta)}
.
\end{gather*}
It turns out that this deformed $AdS_5$ geometry corresponding to the operator $\hat{R}^{{\mathbb{P}}_1}$ has a curvature singularity at $\rho = \infty$ and another singularity at a value of $\rho$ which depends on the angle $\zeta$. When $\zeta = \pi/2$ this singularity is at $\rho = 1/\varkappa$.
The correspondingly $B$-field has the following non-vanishing components
 \begin{gather*}
 B^{{\mathbb{P}}_1}_{\rho\psi_1} =     -  \frac{1}{2\varkappa} \partial_\rho \log{f(\rho,\zeta)},
 \qquad
  B^{{\mathbb{P}}_1}_{\psi_1 \zeta} =    \frac{1}{2\varkappa} \partial_\zeta \log{f(\rho,\zeta)},\\
  B^{\mathbb{P}_1}_{\rho t}=-\varkappa\frac{\rho\sin^2\zeta}{s(\rho,\zeta)}
, \qquad
       B^{\mathbb{P}_1}_{\psi_3\zeta}=\varkappa\frac{\rho^2(1+\rho^2)\sin\zeta\cos\zeta}{s(\rho,\zeta)}
.
 \end{gather*}

For the choice of $R$-matrix $\hat{R}^{{\mathbb{P}}_2}$, the non-zero components of the metric are
\begin{gather*}
G^{{\mathbb{P}}_2}_{tt} = -(1+\rho^2),
\quad
G^{{\mathbb{P}}_2}_{\rho\rho}  = \frac{1+\varkappa ^2+\varkappa ^2 \rho ^2 \left(2+\rho ^2\right) \cos ^2(\zeta
   )}{\left(1+\rho ^2\right) f(\rho ,\zeta ) h(\rho ,\zeta )},
   \qquad
   G^{{\mathbb{P}}_2}_{\zeta \zeta}  =\frac{\rho ^2 \left(1+\varkappa ^2 \left(1+\rho ^2\right)\right)}{f(\rho ,\zeta )
   h(\rho ,\zeta )},\\
     G^{{\mathbb{P}}_2}_{\psi_1\psi_1} = \frac{\rho ^2 \cos ^2{\zeta}}{f(\rho ,\zeta )},
   \qquad
  G^{{\mathbb{P}}_2}_{\psi_2\psi_2} = \frac{\rho ^2 \sin ^2{\zeta }}{h(\rho ,\zeta )},
   \qquad
   G^{{\mathbb{P}}_2}_{\rho\zeta} =-\frac{\kappa ^2 \rho ^3
\sin (2 \zeta )}{2 f(\rho ,\zeta ) h(\rho ,\zeta )}.
\end{gather*}
This deformation of $AdS_5$ associated with the operator $\hat{R}^{{\mathbb{P}}_2}$ has no singularity for finite values of $\rho$. However, the metric
and the curvature scalar diverge when $\rho$ tends to infinity.
The result for the $B$-field is
  \begin{gather*}
 B^{{\mathbb{P}}_2}_{\rho\psi_1} =
  - \frac{1}{2\varkappa} \partial_\rho \log{f(\rho,\zeta)},
 \qquad
 B^{{\mathbb{P}}_2}_{\psi_1 \zeta} =\frac{1}{2\varkappa} \partial_\zeta \log{f(\rho,\zeta)},
\\
B^{{\mathbb{P}}_2}_{\psi_2 \zeta} =   \frac{\varkappa \rho ^2 \left(1+\rho
   ^2\right) \sin (2 \zeta )}{2 h(\rho
   ,\zeta )},
\qquad
B^{{\mathbb{P}}_2}_{\rho\psi_2} = - \varkappa \frac{ \rho \sin ^2{\zeta }}{h(\rho ,\zeta )}.
 \end{gather*}
Finally, let us mention that all three matrices $\hat R$, $\hat{R}^{{\mathbb{P}}_1}$
and $\hat{R}^{{\mathbb{P}}_2}$ may be extended to solutions of the mCYBE equation
on the whole of $\mathfrak{su}(2,2\vert 4)$.

\subsection{Fermionic sector}
The standard choice for the $R$-matrix acting on $\mathfrak{f}^{[1]}$ simply corresponds to
\beqz
\forall M\in\mathfrak{f}^{[1]},\quad R(M)=[J,M], \quad J=\mbox{diag}(i,i,i,i,0,0,0,0).
\eeqz
Because of this very simple form,
and as already noticed in \cite{Arutyunov:2013ega}, one has
\beqz
\forall M\in\mathfrak{f}^{[1]},\quad\forall g\in SU(2,2)\times SU(4),\quad R_g(M)=R(M).
\eeqz
Thus, for this $R$-matrix one simply has to check the invertibility of the operator $1- \eta R\circ  (P_1 - P_3)$ on $\mathfrak{f}^{[1]}$. This is easily shown to hold.

One may be interested to know, however, what happens in the fermionic sector when one chooses another $R$-matrix. Once again, the various cases may be described in terms of permutations $\mathbb{Q}$, but this time of $8$ objects. Since in this paragraph we are only interested in what happens in the fermionic sector, we consider permutations which do not modify the action of $R$ in the bosonic sector. That is to say that we restrict attention to permutations $\mathbb{Q}$ which neither modify the order of the indices ${1,2,3,4}$, nor that of the indices ${5,6,7,8}$. Any such permutation corresponds to a given Dynkin diagram of the Lie superalgebra $\mathfrak{sl}(4\vert 4)$\begin{footnote}{Strictly speaking, the permutations which differ simply by the interchange of the set of indices ${1,2,3,4}$ with ${5,6,7,8}$ correspond to the same Dynkin diagram. However, they should generically be considered as leading to different deformations, because the two blocks ${1,2,3,4}$ and ${5,6,7,8}$ are subject to different reality conditions when restricting to the real form
$\mathfrak{su}(2,2\vert 4)$.}\end{footnote}. Consider, for instance, the permutation
\beqz
{\mathbb{Q}}_1=\left(\begin{array}{cccccccc}1&2&3&4&5&6&7&8\cr 1&2&5&6&7&8&3&4\end{array}\right),
\eeqz
corresponding to the Dynkin diagram
\begin{equation*}
\begin{tikzpicture}[baseline =-5,scale=.6]
\draw[thick] (1,0) -- (7,0);
\foreach \x in {1,2,3,4,5,6,7}
\filldraw[fill=white, thick] (\x,0) circle (2mm);
\draw[thick] (2,0)++(-.15,-.15) -- ++(.3,.3);\draw[thick] (2,0)++(.15,-.15) -- ++(-.3,.3);
\draw[thick] (6,0)++(-.15,-.15) -- ++(.3,.3);\draw[thick] (6,0)++(.15,-.15) -- ++(-.3,.3);
\end{tikzpicture}
\end{equation*}
The corresponding operator $R^{{\mathbb{Q}}_1}$ has a simple restriction to $\mathfrak{f}^{[1]}$ which reads
 \beqz
 \forall M\in\mathfrak{f}^{[1]},\quad R^{{\mathbb{Q}}_1}(M)=[J_1,M],\quad J_1=\mbox{diag}(i,i,-i,-i,0,0,0,0).
 \eeqz
Because the matrix $J_1$ does not commute with $SU(2,2)$, the operator $R^{{\mathbb{Q}}_1}_g$ depends on $g$. One finds that the restriction to $\mathfrak{f}^{[1]}$ of the operator $1- \eta R^{{\mathbb{Q}}_1}_g \circ  (P_1 - P_3)$ is singular for $\rho = 1/\varkappa$. For comparison, let us consider another possible permutation
\beqz
\mathbb{Q}_2=\left(\begin{array}{cccccccc}1&2&3&4&5&6&7&8\cr 5&6&1&2&3&4&7&8\end{array}\right),
\eeqz
which corresponds to the same Dynkin diagram. The restriction of $R^{{\mathbb{Q}}_2}$ to $\mathfrak{f}^{[1]}$ again has a simple form
 \beqz
 \forall M\in\mathfrak{f}^{[1]},\quad R^{{\mathbb{Q}}_2}(M)=[J_2,M],\quad J_2=\mbox{diag}(0,0,0,0,i,i,-i,-i).
 \eeqz
In this case, one finds that the restriction to $\mathfrak{f}^{[1]}$ of the operator $1- \eta R^{{\mathbb{Q}}_2}_g \circ  (P_1 - P_3)$ is regular for finite values of $\rho$.
Yet another example of a permutation is
\beqz
\mathbb{Q}_3=\left(\begin{array}{cccccccc}1&2&3&4&5&6&7&8\cr 1&5&6&7&2&3&4&8\end{array}\right),
\eeqz
corresponding to the Dynkin diagram
\begin{equation*}
\begin{tikzpicture}[baseline =-5,scale=.6]
\draw[thick] (1,0) -- (7,0);
\foreach \x in {1,2,3,4,5,6,7}
\filldraw[fill=white, thick] (\x,0) circle (2mm);
\draw[thick] (1,0)++(-.15,-.15) -- ++(.3,.3);\draw[thick] (1,0)++(.15,-.15) -- ++(-.3,.3);
\draw[thick] (4,0)++(-.15,-.15) -- ++(.3,.3);\draw[thick] (4,0)++(.15,-.15) -- ++(-.3,.3);
\draw[thick] (7,0)++(-.15,-.15) -- ++(.3,.3);\draw[thick] (7,0)++(.15,-.15) -- ++(-.3,.3);
\end{tikzpicture}
\end{equation*}
The restriction of $R^{{\mathbb{Q}}_3}$ to $\mathfrak{f}^{[1]}$ cannot be written as a commutator. Nevertheless, one can show that the restriction to $\mathfrak{f}^{[1]}$ of the operator $1- \eta R^{{\mathbb{Q}}_3}_g \circ  (P_1 -P_3)$ is regular for finite values of $\rho$.

\providecommand{\href}[2]{#2}\begingroup\raggedright\endgroup

\end{document}